\documentclass[aps,pra,twocolumn,floatfix,superscriptaddress]{revtex4-2}
\usepackage[utf8]{inputenc}
\usepackage[english]{babel}

\usepackage{graphicx}
\usepackage{amsmath,times}
\usepackage{amsfonts}
\usepackage{mathrsfs}
\usepackage{physics}
\usepackage[colorlinks=true,hyperfootnotes=true,breaklinks=true,citecolor=red,urlcolor=blue,linkcolor=blue]{hyperref}
\usepackage{mathbbol} 

\usepackage{soul}

\begin{document}

\title{Coin dimensionality as a resource in quantum metrology involving discrete-time quantum walks}

\author{Simone Cavazzoni}
\email{simone.cavazzoni@unimore.it}
\affiliation{Dipartimento di Scienze Fisiche, Informatiche e Matematiche, 
Universit\`{a} di Modena e Reggio Emilia, I-41125 Modena, Italy}

\author{Luca Razzoli}
\email{luca.razzoli@uninsubria.it}
\affiliation{Center for Nonlinear and Complex Systems, Dipartimento di Scienza e Alta Tecnologia, 
Universit\`a degli Studi dell’Insubria, I-22100 Como, Italy}
\affiliation{INFN, Sezione di Milano, I-20133 Milano, Italy}

\author{Giovanni Ragazzi}
\email{giovanni.ragazzi@unimore.it}
\affiliation{Dipartimento di Scienze Fisiche, Informatiche e Matematiche, 
Universit\`{a} di Modena e Reggio Emilia, I-41125 Modena, Italy}

\author{Paolo Bordone}
\email{paolo.bordone@unimore.it}
\affiliation{Dipartimento di Scienze Fisiche, Informatiche e Matematiche, 
Universit\`{a} di Modena e Reggio Emilia, I-41125 Modena, Italy}
\affiliation{Centro S3, CNR-Istituto di Nanoscienze, I-41125 Modena, Italy}

\author{Matteo G. A. Paris}
\email{matteo.paris@fisica.unimi.it}
\affiliation{Quantum Technology Lab, Dipartimento di Fisica {\em Aldo Pontremoli}, Universit\`{a} degli Studi di Milano, I-20133 Milano, Italy}
\affiliation{INFN, Sezione di Milano, I-20133 Milano, Italy}

\date{\today}

\begin{abstract}
We address metrological problems where the parameter of interest is encoded in the internal degree of freedom of a discrete-time quantum walker, and provide evidence that coin dimensionality is a potential resource to enhance precision. In particular, we consider estimation problems where the coin parameter governs rotations around a given axis and show that the corresponding quantum Fisher information (QFI) may increase with the dimension of the coin. We determine the optimal initial state of the walker to maximize the QFI and discuss whether, and to which extent, precision enhancement may be achieved by measuring only the position of the walker. Finally, we consider Grover-like encoding of the parameter and compare results with those obtained from rotation encoding. 
\end{abstract}
\maketitle

\section{Introduction}
\label{Introduction}
Discrete-time quantum walks (DTQWs) are quantum generalizations of classical 
random walks describing the discrete time propagation of a quantum particle 
over a discrete space \cite{wang2013physical,portugal2018quantum,KADIAN2021100419}. 
In a DTQW the motion of the walker is determined at each time step by the state of its internal degree of freedom, the so-called \textit{coin}. Entanglement thus arises between  the spatial degree of freedom of the particle and its "spin" (or polarization). 
DTQW models have found applications in quantum search algorithms 
\cite{PhysRevA.67.052307} and provide a mean for universal quantum computation \cite{ambainis2003quantum,lovett2010universal,singh2021universal} and simulation of quantum phenomena \cite{chatterjee2020experimental}. More recently, DTQWs has been also suggested for image and data encryption \cite{abd2021optical,abd2020secure}, and for link prediction \cite{liang2022hadamard}. DTQWs have been implemented in a
wide range of physical systems including optical-quantum systems \cite{travaglione2002implementing,dur2002quantum,karski2009quantum,ehrhardt2021exploring}, wave-guide lattices \cite{perets2008realization,broome2010discrete}, silicon based systems \cite{qiang2021implementing}, spin systems \cite{tang2020experimental}, spin orbit photonics \cite{di2023ultra}, and Bose Einstein condensate \cite{xie2020topological} or lattices \cite{esposito2022quantum}, as well as in quantum computers \cite{acasiete2020implementation}.

The dynamics of DTQWs is described by a unitary operator which governs the evolution of the whole system, particle plus coin, in a single time step. It consists of a unitary operation applied on the coin state, followed by a unitary {\em conditional} shift operator which makes the walker change position depending on the coin state. In the prototypical 
DTQW, a walker propagating on a line, it may only hop to the left or to the right 
depending on the state of its two-dimensional, $D=2$, internal degree of freedom. This basic principle can be extended to generate a rich variety of DTQW, with methods 
that range from changing the coin operator at each time step \cite{PhysRevA.80.042332,PhysRevE.104.064209,PhysRevLett.111.180503} to applying multiple coin operators at the same time step, and by generalizing DTQWs to generic graphs and higher dimensional coin spaces \cite{kendon2006,lovett2010universal,PhysRevA.92.040302,PhysRevResearch.2.023378}, e.g, 
by coupling the walker to more than one coin 
\cite{PhysRevA.67.052317,segawa2008,shang2018} or to a single coin of higher 
dimension, $D > 2$ \cite{mackay2002,Hamilton_2011,PhysRevA.90.012307}. 

In this work, we focus on the latter approach, and refer to a DTQW with a $D$-dimensional coin as a $D$-state walk. For even $D = 2k$, with $k\in \mathbb{N}$, at each time step the walker is allowed to hop up to its $k$th nearest neighbouring site, but is not allowed to stay on its current position \cite{lorz2019photonic}, a possibility granted only for odd $D = 2k +1$ \cite{stefanak2012continuous}. Increasing $D>2$ paves the way to explore dynamics that are not possible in a usual two-state walk. In the simplest examples, i.e.,  
passing from $D=2$ to $D=3$ the walker is also allowed to stay at its current position \cite{stefanak2012continuous}, while for $D=4$ it is allowed to reach the next-nearest neighbours, but not to stay on the current site \cite{PhysRevLett.97.023602,lorz2019photonic}.

The metrological interest of DTQWs lies in the entanglement that is established between the two degrees of freedom (walker's position and coin), making the overall system more sensitive to tiny variations of the parameter, or making one of them available to probe the other. In particular, we focus on metrological problems where the parameter of interest is encoded in the internal degree of freedom of the walker \cite{annabestani2022multiparameter}, and provide evidence that coin dimensionality is a potential resource to enhance precision. We determine the optimal initial state of the walker to maximize the quantum Fisher information, and discuss whether and to which extent, precision enhancement may be achieved by measuring only the position of the walker \cite{singh2019quantum}. 
Indeed, coin dimension plays a fundamental role in the time evolution of the DTQW, also in the position space and, in turn, in controlling and engineering the system. 
Our results may thus find applications in the characterization of DTQW-based quantum computing \cite{PhysRevA.77.032326,inproceedings,Chawla2023}, as well as in estimation of node proximity \cite{wang2021qsim} and in specific scattering problems \cite{zatelli2020scattering}. \textcolor{black}{The relation between metrology and coin dimensionality can be exploited for the study of polarized photons or particles with spin, where the coin plays the role of the internal degree of freedom (i.e. the polarization or the spin) of the system \cite{lorz2019photonic,tang2020experimental,mallick2020spectral}}.

The paper is organized as follows: in Section \ref{DTQW} we introduce the DTQW model and the different encodings---single-parameter coin operators---we will consider for the estimation problem. In Section \ref{QFI} we briefly review the concepts of classical and quantum Fisher information and we state the metrological problem considered in the present work. In Sections \ref{Main_Results}--\ref{s:gr} we present analytical and numerical results of the estimation problem for different encoding of the coin parameter and dimension of the coin. Section \ref{conclusion} is devoted to the conclusions and perspectives. Further details and proofs can be found in the Appendices.

\section{Discrete Time Quantum Walks}
\label{DTQW}
One-dimensional DTQW models describe the time evolution of a quantum particle 
(walker) with an internal degree of freedom (coin), say spin in the following, over an infinite, discrete line. The Hilbert space of this bipartite system walker+coin is $\mathscr{H} = \mathscr{H}_{p} \otimes \mathscr{H}_{c}$, with $\mathscr{H}_p = \operatorname{span}\{\vert x \rangle_p \,\vert\, x \in \mathbb{Z}\}$ the position space, and $\mathscr{H}_{c} = \operatorname{span}\{\vert m \rangle_c \,\vert\, m \in I_c^{(s)} \}$ the coin space, where  $\dim \mathscr{H}_c = D = 2s+1$ for a spin-$s$  particle. Note that (half-)integer $s$ corresponds to (even) odd $D$. For later convenience, we have introduced the set of integer indices
\begin{equation}
    I_c^{(s)} =
    \begin{cases}
     \{ -s,\ldots,-1,0,1, \ldots, s \} & \text{(integer $s$)},\\
	\{ -s-\frac{1}{2},\ldots,-1,1, \ldots, s+\frac{1}{2} \} & \text{(half-integer $s$)},
    \end{cases}
    \label{eq:set_indices_spin_coin}
\end{equation}
sorted in ascending order and relating, in a one-to-one correspondence, the quantum number $-s \leq m_s \leq s$ associated to the $z$-axis component of the spin $s$, which can be half-integer, to the shift in position space of the walker, which is an integer, via
\begin{equation}
\begin{array}{ccccccccl}
m_s&\in	&\{ &-s,	& -s+1,	& \ldots, & s-1,		& s &\}\\
\updownarrow\\
m&\in	 	&\{ & i_1,	& i_2,	& \ldots, & i_{D-1},	& i_D& \}.
\end{array}
\end{equation}
The indices $i_k \in I_c^{(s)}$ satisfy the relation $i_{k+1} = i_k+1$, with the only exception of the index $i_k=0$ which is not included in the set associated to half-integer $s$.

The evolution of the DTQW for one time step is defined by the unitary operator
\begin{equation}
\label{time_evolution_operator}
\mathcal{U} = \mathcal{S} \left( \mathbb{1}_p \otimes \mathcal{C}  \right),
\end{equation}
with $\mathbb{1}_p$ the identity in the position space $\mathscr{H}_p$. At a given time step, the coin state is changed by applying the coin operator $\mathcal{C}$, leaving the walker's position state unaltered, and this operation is followed by a conditional shift operator,
\begin{equation}
\label{eq:conditional_shift_operator}
\mathcal{S} = \sum_{x \in \mathbb{Z}} \sum_{m \in I_c^{(s)}} \vert x + m \rangle_p {}_p\langle x \vert \otimes \vert m \rangle_c{}_c\langle m \vert
\end{equation}
with $I_c^{(s)}$ defined in Eq. \eqref{eq:set_indices_spin_coin}, which makes the walker evolve in position space according to the coin state $\vert m \rangle_c$. Both the operators $\mathcal{C}$ and $\mathcal{S}$ must be unitary for $\mathcal{U}$ to be unitary.
Assuming a single, constant and uniform coin operator,\footnote{I.e., a coin operator 
which does depend neither on time nor on the position of the walker.} the state of the system at time $t \in \mathbb{N}$ is thus
\begin{equation}
\vert \psi(t) \rangle = \mathcal{U} \vert \psi \left( t-1 \right) \rangle  = \mathcal{U}^{t} \vert \psi ( 0 ) \rangle,
\end{equation}
with $\vert \psi (0) \rangle$ the initial state of the system. 

Our aim is to investigate estimation problems where the parameter $\theta$ to be estimated is encoded in the coin operator and to determine the optimal probe for this purpose, assuming an initially localized walker. In the following Sections, we introduce the 
single-parameter coin operators and the probe considered in the present work.

\subsection{Coin operators}
\label{sec:coin_ops}
We describe now three different ways of encoding the parameter of interest in the DTQW dynamics, i.e., the three types of single-parameter coin operators $\mathcal{C}$ investigated in the present work. 
As a first coin model, we consider the operator (we set $\hbar = 1$)
\begin{equation}
\label{generators}
R^{(D)} _{\hat{n}} (\theta) = e^{-i \theta_{\hat{n}} \cdot \mathcal{T}^{(D)}_{\hat{n}}},
\end{equation}
which rotates the spin $s$ of an angle $\theta$ about the axis $\hat{n}=\hat{x},\hat{y},\hat{z}$ (unit vector). The generators $\mathcal{T}^{(D)}_{\hat{n}}$ of the $(D=2s+1)$-dimensional rotation (see Appendix \ref{app:spin_rotation_gen} for their matrix representation) satisfy the relation
\begin{equation}
\label{lie}
\left[ \mathcal{T}^{(D)}_{a}, \mathcal{T}^{(D)}_{b} \right] = i \epsilon_{abc} \mathcal{T}^{(D)}_{c}
\end{equation}
where $ \epsilon_{abc} $ is the Levi-Civita symbol.

As a second coin model, we embed the most general form of the coin operator in $D=2$ \textcolor{black}{(obtained through the Euler angles parametrization \cite{cacciatori2022compact} and neglecting an overall phase factor), i.e., an element of $U(2)$,
\begin{equation}
\label{eq:U2coin_general}
\mathcal{C}_{\xi,\theta,\zeta}^{({D}=2)} = 
\begin{pmatrix}
e^{-i \frac{\xi+\zeta}{2}}\cos \frac{\theta}{2} & -e^{i \frac{\xi-\zeta}{2}}\sin \frac{\theta}{2}\\
e^{-i \frac{\xi-\zeta}{2}}\sin \frac{\theta}{2} & e^{i \frac{\xi+\zeta}{2}}\cos \frac{\theta}{2} \\
\end{pmatrix}
\end{equation}
where $\xi \in [0,4\pi],\theta \in [0,\pi],\zeta \in [0,2\pi]$, into a higher dimensional space as
\begin{equation}
\mathcal{C}_{\xi,\theta,\zeta}^{(E,D>2)} = 
\begin{pmatrix}
e^{-i \frac{\xi+\zeta}{2}}\cos \frac{\theta}{2} & 0 & \dots & 0 & -e^{i \frac{\xi-\zeta}{2}}\sin \frac{\theta}{2}  \\
0 & 1 & 0 & \dots & 0 \\
\vdots & \ddots & \ddots & \ddots & \vdots \\
0 & 0 & \ddots & 1 & 0 \\
e^{-i \frac{\xi-\zeta}{2}}\sin \frac{\theta}{2}  & 0 & \dots & 0 & e^{i \frac{\xi+\zeta}{2}}\cos \frac{\theta}{2} 
\end{pmatrix} .
\label{eq:2D_coin_embedded}
\end{equation}}
The effective two-dimensional coin operator acts on the coin states with the lowest and highest index, while leaving the others unaffected (identity). In dimension $D$ there are $D(D-1)/2$ independent embeddings, but we focus on this one due to its relation to Euler angles passing from $D=2$ to $D=3$ \cite{byrd1997geometry}.

As a third coin model, we consider the so-called \textit{generalized Grover coin}, \textcolor{black}{that for $D=2,3$, is defined as
\begin{equation}
\label{2D_grover}
\mathcal{C}_{G}^{(2)}(\theta)=\begin{pmatrix}
\theta & \sqrt{1-\theta^{2}} \\
\sqrt{1-\theta^{2}} & -\theta
\end{pmatrix} ,
\end{equation}
\begin{equation}
\label{3D_grover}
\mathcal{C}_{G}^{(3)}(\theta)=\begin{pmatrix}
-\theta^{2} & \theta\sqrt{2-2\theta^2} & 1-\theta^{2} \\
\theta\sqrt{2-2\theta^2} & 2\theta^{2}-1 & \theta\sqrt{2-2\theta^2} \\
1-\theta^{2} & \theta\sqrt{2-2\theta^2} & -\theta^{2}
 \end{pmatrix},
\end{equation}
with the coin parameter $0 \leq \theta \leq 1$. Note that $\mathcal{C}_{G}^{(2)}(\theta=1/\sqrt{2})$ is the Hadamard coin ($D=2$) \cite{endo2020eigenvalues,liang2022hadamard} and $\mathcal{C}_{G}^{(3)}(\theta=1/\sqrt{3})$ is the Grover coin in $D=3$ \cite{stefanak2012continuous,li2015one}.} The Grover coin was named after showing that a DTQW can be used to implement Grover’s search algorithm using Grover’s diffusion operator on the coin space \cite{moore2002proceedings,PhysRevA.67.052307,tregenna2003controlling}. Then, attempts have been made to generalize it by considering parametric coin operators which recover the Grover coin for some specific values of the parameters \cite{stefanak2012continuous,stefanak2014limit,sarkar2020periodicity,mandal2022limit}. Accordingly, DTQWs with a generalized Grover coin are generally referred to as generalized Grover walks. 

\subsection{The initial state}
As a probe, we consider pure separable initial states 
\begin{equation}
\vert \psi (0) \rangle   = \vert 0 \rangle_p \otimes \vert \phi^{(D)} \rangle_{c},
\label{eq:probe_state}
\end{equation}
where we assume that the walker is initially localized at the origin of the line, $x=0$,\footnote{All the points are equivalent in the infinite line, so our assumption is just to have an initially localized walker.} i.e., an eigenstate of the position operator, $X \vert x \rangle_p = x \vert x \rangle_p$. Our purpose is to determine the optimal preparation of the $D$-dimensional coin state $\vert \phi^{(D)} \rangle_c \in \mathscr{H}_c$ for estimating the parameter encoded in the coin operator.

A coin pure state $\vert \phi^{(D)} \rangle_c \in \mathscr{H}_c$, with $\dim \mathscr{H}_c = D $, is in principle identified by $D$ complex coefficients $\{\chi_m\}_m$ and can be written as
\begin{equation}
\label{eq:arbitrary_coin_state}
    \vert \phi^{(D)} \rangle_c = \sum_{m \in I_c^{(s)}} \chi_m \vert m \rangle_c.
\end{equation}
The actual number of independent real parameters is reduced to $2({D}-1)$ parameters by the normalization condition and because of the overall arbitrary phase, which is physically meaningless \cite{mendas2008}. Here we parameterize the coin state using ${D}-1$ angles $\alpha_i$, as a portion of a $D$-dimensional surface of an hypersphere embedded in a $({D}+1)$-dimensional space. The configurational space is obtained through the rotation of this portion of hypersuface by ${D}-1$ phases $\gamma_i$. These parameters can take values

\begin{equation}
\label{parameters}
\begin{cases}
0 \leq \alpha_{i} \leq \pi ,\\
0 \leq \gamma_{i} \leq 2 \pi ,
\end{cases}
\end{equation}
with $i=1,\ldots,D-1$. This parametrization generalizes the Bloch sphere (recovered in dimension $D=2$) to higher dimension $D>2$. Accordingly,  a generic state is parametrized as
\begin{align}
    \label{D_Bloch_Sphere}
    &\vert \phi^{(D)} \rangle_c  =  \cos \left( \frac{\alpha_{1}}{2} \right) \vert i_1 \rangle_c + e^{i\gamma_{1}} \prod_{j=1}^{D-1}\sin\left( \frac{\alpha_{j}}{2} \right) \vert i_{2} \rangle_c \nonumber \\  
    & + \sum_{k=2}^{D-1} e^{i\gamma_{k}} \left( \prod_{j=1}^{D-k}\sin\left( \frac{\alpha_{j}}{2} \right) \right) \cos\left( \frac{\alpha_{D+1-k}}{2} \right) \vert i_{1+k} \rangle_c,
\end{align}
where $i_k \in I_{c}^{(s)}$ in Eq. \eqref{eq:set_indices_spin_coin}.

In this work we focus on coins of dimension $D=2,3,4$, so the generic state $\vert \phi^{(D)} \rangle_c$ can be explicitly written as
\begin{align}
\vert \phi^{(2)} \rangle_c & =  \cos \left( \frac{\alpha_{1}}{2} \right) \vert -1 \rangle_c + e^{i\gamma_{1}} \sin \left( \frac{\alpha_{1}}{2} \right) \vert +1 \rangle_c,
\label{Bloch_Sphere}\\
\vert \phi^{(3)} \rangle_c & =   \cos\left( \frac{\alpha_{1}}{2} \right) \vert -1 \rangle_c\nonumber + e^{i\gamma_{1}} \sin\left( \frac{\alpha_{1}}{2} \right) \sin\left( \frac{\alpha_{2}}{2} \right)\vert 0 \rangle_c \\
& + e^{i\gamma_{2}}\sin\left( \frac{\alpha_{1}}{2} \right) \cos\left( \frac{\alpha_{2}}{2} \right) \vert +1 \rangle_c
,\label{3D_Bloch_Sphere}\\
\vert \phi^{(4)} \rangle_c & = \cos\left( \frac{\alpha_{1}}{2} \right)  \vert -2 \rangle_c  \nonumber \\
& + e^{i\gamma_{1}} \sin\left( \frac{\alpha_{1}}{2} \right) \sin\left( \frac{\alpha_{2}}{2} \right) \sin\left( \frac{\alpha_{3}}{2} \right) \vert -1 \rangle_c \nonumber \\  
& + e^{i\gamma_{2}} \sin\left( \frac{\alpha_{1}}{2} \right) \sin\left( \frac{\alpha_{2}}{2} \right) \cos\left( \frac{\alpha_{3}}{2} \right) \vert +1 \rangle_c \nonumber \\
& + e^{i\gamma_{3}}\sin\left( \frac{\alpha_{1}}{2} \right) \cos\left( \frac{\alpha_{2}}{2} \right) \vert +2 \rangle_c \label{4D_Bloch_Sphere}.  
\end{align}
The coin, being a $D$-level system, can be thought of as a qudit. We chose the above parametrization of the coin state for consistency with such an interpretation, qubit state  \eqref{Bloch_Sphere}, qutrit state \eqref{3D_Bloch_Sphere}  \cite{CAVES2000439}, up to $D=4$ in Eq. \eqref{4D_Bloch_Sphere}.

\section{Quantum metrology in DTQWs}
\label{QFI}
In our framework, the coin operator $\mathcal{C}$ depends on a single unknown parameter $\theta$. The DTQW, intended as the time evolution of the system, strongly depends on such parameter. Our purpose is to investigate the estimation problem for such parameter with emphasis on the effects of coin dimensionality.

\subsection{Classical and quantum Fisher information}
Given an observable $X$ with outcomes $\{x\}$ characterized by the conditional probability $p(x\vert \theta)$ of obtaining the value $x$ when the parameter takes the value $\theta$, the Fisher information (FI)
\begin{equation}
\label{general_fi}
F_{X}(\theta) = \int dx \frac{\left[ \partial_{\theta} p( x \vert \theta ) \right]^{2}}{p\left(x \vert \theta \right)} ,
\end{equation}
provides a measure of the amount of information that the observable $X$ carries about the parameter $\theta$.

In fact, for unbiased estimators, upon performing $M$ measurements of the observable $X$, the variance of the parameter $\theta$ to be estimated satisfies the Cram\'{e}r-Rao inequality
\begin{equation}
\label{Cramer_Rao}
\operatorname{Var}\left( \theta \right) \geq \dfrac{1}{M F_{X}\left( \theta \right)}.
\end{equation}
In our case, the relevant observable is the position of the walker, which has 
a discrete spectrum, outcomes $x \in \mathbb{Z}$, hence the FI is
\begin{equation}
F_{X}(\theta) = \sum_{x \in \mathbb{Z}} \frac{\left[ \partial_{\theta} p( x \vert \theta ) \right]^{2}}{p\left(x \vert \theta \right)}.
\label{fisher_information}
\end{equation}
In particular, our quantum system is bipartite (walker's position + coin), but we would perform a projective measurement on a part of it (walker's position). Therefore, if we denote by $\rho_\theta$ the density matrix of the bipartite system parametrized  by $\theta$, the conditional probability distribution we need to assess the FI is
\begin{equation}
     p\left( x \vert \theta \right) = \langle x \vert \Tr_{c}\left[\rho_\theta\right] \vert x \rangle,
\end{equation}
i.e., it follows from projecting the reduced density matrix of the walker (obtained from a partial trace of the total density matrix over the coin space) onto the position state $\vert x \rangle$.

Moving to the quantum realm, the parameter is encoded in the state of the quantum system, $\rho_\theta$, and the figure of merit to consider is the quantum Fisher information (QFI) \cite{paris2009quantum}
\begin{equation}
H(\theta) = \Tr[\rho_{\theta}\mathcal{L}_{\theta}^2] \geq F_X(\theta)\quad \forall X,
\label{eq:QFI_def_geq_FI}
\end{equation}
that is independent of the selected measurement procedure, and with $\mathcal{L}$ 
the symmetric logarithmic derivative defined by the implicit relation
\begin{equation}
\label{sld}
\partial_{\theta} \rho_{\theta} = \frac{1}{2} (\mathcal{L}_{\theta}\rho_{\theta} + \rho_{\theta}\mathcal{L}_{\theta}).
\end{equation}
The FI is always bounded from above by the QFI for any quantum measurement, thus the quantum Cram\'{e}r-Rao inequality
\begin{equation}
\label{Quantum_Cramer_Rao}
\operatorname{Var}\left( \theta \right) \geq \dfrac{1}{M H(\theta )}
\end{equation} 
sets the ultimate lower (quantum) bound on the achievable precision in estimating the parameter $\theta$.
As a final remark, we point out that for pure states $\vert \psi_\theta \rangle$, 
as those of the overall DTQW, the QFI reduces to \cite{paris2009quantum}
\begin{equation}
\label{quantum_fisher_information}
H(\theta) = 4 \left( \langle \partial_{\theta} \psi_{\theta} \vert \partial_{\theta} \psi_{\theta} \rangle - \left| \langle \partial_{\theta} \psi_{\theta} \vert \psi_{\theta} \rangle\right|^2 \right).
\end{equation}

\subsection{The metrological problem}
The metrological problem we address in this work concerns the estimation of the parameter encoded in a given coin operator when letting the probe state \eqref{eq:probe_state} evolve in time, performing a DTQW. The coin operators considered are the $x,y,z$-rotations and the generalized Grover coin in dimension $D=2,3,4$ (see Sec. \ref{sec:coin_ops}). For the different coins, we discuss the dependence of the QFI on time and on the dimension of the coin, and determine the optimal preparation of the probe, $\vert \Phi^{(D)} \rangle_c$, to maximize the QFI. In addition, we compare the latter with the FI associated to a position measurement of the walker, which is the natural measurement in a QW. We recall that a measurement is said to be optimal if the corresponding FI saturates the bound in Eq. \eqref{eq:QFI_def_geq_FI}, i.e., whenever the FI equals the QFI. Unless otherwise specified, the optimal probes have been either numerically determined or analytically induced, following Appendix \ref{app:Analytical Calculations}, and numerically verified. In the following sections we present results for rotations encodings (Sec. \ref{Main_Results}--\ref{s:xy}) and for the generalized Grover coin encoding (Sec. \ref{s:gr}).

\section{Metrology with $z$-rotations encoding and different coin dimensions}
\label{Main_Results}
The $z$-rotation operator is diagonal in the chosen basis for the coin space, because the coin basis states are eigenstates of it. This allows us to analytically determine the evolution of the DTQW, and so the FI and the QFI. We start by discussing the results for any $D$ (rotations and embedded rotations) and then we refine the discussion for $D=2,3,4$ 
as case studies.

\subsection{Results for arbitrary dimension $D$}
\label{sec:QFI_RzD}

\subsubsection{Actual rotation}
The $z$-rotation in arbitrary $D$-dimensional coin space,
\begin{equation}
\label{rzD}
R^{(D)}_{z}(\theta) = 
\operatorname{diag}\left( \{e^{i m_s \theta}\}_{-s \leq m_s \leq s}\right),
\end{equation}
is a diagonal matrix where the index $m_s$ is the quantum number associated to the $z$-component of the spin $s = (D-1)/2$. The initial state of the system (see Eq.\eqref{eq:probe_state}), with initial coin state as in Eq. \eqref{eq:arbitrary_coin_state}, evolves according to
\begin{equation}
    \vert \psi(t) \rangle = \mathcal{U}^t \vert \psi(0) \rangle
    =  \sum_{m \in I_c^{(s)}} e^{-i \theta m_s t} \chi_{m} \vert m t\rangle_p \otimes  \vert m \rangle_c,
    \label{eq:psit_RzD}
\end{equation}
where the index of summation has a twofold role: it runs over the quantum number $-s \leq m_s \leq s$ and, correspondingly, over the associated integer shift $m$ (see Eq. \eqref{eq:set_indices_spin_coin}). The corresponding QFI \eqref{quantum_fisher_information} is
\begin{equation}
    H^{(D)}_{z}(t) = 4 t^2 \! \bigg[ \sum_{m \in I_c^{(s)}} \!\!\!\! m_s^2 \vert \chi_{m} \vert^2 - \bigg( \sum_{m \in I_c^{(s)}} m_s \vert \chi_{m} \vert^2 \bigg)^2 \bigg] 
    \label{eq:QFI_arbitrary_D_RzD}
\end{equation}
and depends neither on $\theta$  (the parameter to be estimated) nor on the phases $\{\gamma_i\}_i$ of the probe state. It only depends on the angles $\{ \alpha_i\}_i$ of the latter via $\{\vert \chi_m \vert^2\}_m$.

The QFI is minimum (null) when the probe state is a basis state (eigenstate of the $z$-rotation operator), $\chi_m = \delta_{m,m'}$. In that case, the system evolves as $\vert \psi(t) \rangle = e^{-i \theta m_s' t} \vert m' t\rangle_p \otimes  \vert m' \rangle_c$, i.e., it gains an overall phase factor, the coin state is unchanged, and, at time $t$, the walker is localized at $x_t = m't$. In particular, the walker remains at the origin if $m'=0$ and performs jumps of amplitude $x_{t+1}-x_{t} = m'$ at each time step. The global phase factor, which is the only term still encoding $\theta$, has no physical meaning and indeed it provides null QFI.

Our purpose is to maximize the QFI with respect to the initial coin state, i.e., to the coefficients $\{ \chi_{m} \}_m$ that satisfy $\sum_{m \in I_c^{(s)}} \vert \chi_{m} \vert^2 = 1$. Because of the normalization constraint, we can maximize the QFI by  weighting only the coefficients $\chi_{m}$ with the largest $ m_s^2$, that is $m_s = \pm s$, thus assuming $\chi_{m}=0$ if $\vert m_s \vert \neq s$. We denote by $\pm M$ the integer indices $m \in I_c^{(s)}$ respectively associated to $m_s = \pm s$, with
\begin{equation}
    M =
    \begin{cases}
     s       & \text{if $s$ is integer (odd $D$)},\\
     s+1/2 & \text{if $s$ is half-integer (even $D$)},
    \end{cases}
    \label{eq:index_M_Rz}
\end{equation}
from Eq. \eqref{eq:set_indices_spin_coin}. Accordingly, the QFI simplifies to
\begin{equation}
    H^{(D)}_{z}(t) = 4 t^2 s^2\left[ 1 - (\vert \chi_{M} \vert^2 -\vert \chi_{-M} \vert^2)^2\right],
\end{equation}
which is maximum for $\vert \chi_{M} \vert = \vert \chi_{-M} \vert$. Recalling that ${D}=2s+1$, the maximum QFI achievable in dimension ${D}$ or for a spin-$s$ particle is
\begin{equation}
\label{max_QFI_Rz}
\max_{\vert \phi^{(D)} \rangle_c} H^{(D)}_{z}(t) = \left( {D}-1 \right)^{2} t^2 = 4 s^2 t^2.
\end{equation}
The QFI is quadratic in $D$, so, in this sense, a higher dimension of the coin is a metrological resource for estimating the parameter $\theta$ encoded in the coin operator. The corresponding optimal initial coin state is
\begin{equation}
    \label{initial_max_QFI_Rz}
    \vert \Phi_{z}^{(D)} \rangle_c = \frac{1}{\sqrt{2}} \left( \vert -M \rangle_c + e^{i \gamma} \vert M \rangle_c \right) .
\end{equation}
This expression follows from $\vert \chi_{-M} \vert = \vert \chi_{+M} \vert$, the normalization condition, and the fact that we can neglect an overall global phase factor. An alternative proof of the maximization of the QFI and the optimal probe is offered in Appendix \ref{app:alt_proof_QFI_max_RzD}.

We now focus on the FI of measuring the walker's position. After performing a partial trace over the coin's degrees of freedom, the reduced density matrix resulting from the state \eqref{eq:psit_RzD} is
\begin{equation}
    \rho_p(t) = \sum_{m \in I_c^{(s)}} \vert \chi_{m} \vert^2 \vert m t \rangle_p {}_p \langle  m t  \vert,
    \label{eq:redrho_p_RzD}
\end{equation}
diagonal in position space, with probabilities that are independent of $\theta$ and $t$. Therefore, for the coin  $R_{z}^{(D)}(\theta)$ we have $F_X(\theta) \equiv 0$ for any $D$, i.e., we cannot gain any information on the parameter $\theta$ by measuring the walker's position. Note that, being $\rho_p$ diagonal in the position space, the coherence of the reduced density matrix is null. Analogously, the FI of a momentum measurement is identically null, as the probability distribution does not depend on $\theta$ in position space and thus neither in momentum space after performing a Fourier transform.

One may wonder whether entanglement between the walker and the coin plays any role in the estimation problem. To address this question, we consider the von Neumann entropy $\mathcal{E} = -\Tr(\rho_{p} \log \rho_{p})$, with $\rho_{p}$ the reduced density matrix of the walker's position (equivalently with the coin's reduced density matrix). The bounds are $0 \leq \mathcal{E} \leq \log D$, with $D$ the dimension of the lower-dimensional subsystem between the two, here the coin. The general reduced density matrix \eqref{eq:redrho_p_RzD} is diagonal and its eigenvalues, $\vert \chi_m \vert^2$, do not depend on time. Assuming the optimal probe \eqref{initial_max_QFI_Rz} as the initial state and focusing on $t>0$ (the initial state is separable, so $\mathcal{E}(t=0)=0$), $\rho_p$ admits only two nonzero eigenvalues, $\vert \chi_{\pm M} \vert^2=1/2$, and so $\mathcal{E} = \log 2$. This result reveals the following: (i) The degree of entanglement $\mathcal{E}=\log 2$ is constant in time for $t>0$, (ii) it is independent of $D$, and (iii) it is maximum only in $D=2$. The optimal probe for the estimation problem generates entanglement between walker's position and coin, but it is not maximum in $D>2$. Therefore, in the following we will not further investigate entanglement for other encodings, as this example already shows that, at least for an initially localized walker, in general we can not expect the optimal probe to generate maximal entanglement, i.e., we cannot expect to have a direct implication between maximum QFI and maximum entanglement.

\subsubsection{Embedding in dimension $D>2$}
\label{sec:QFI_REzD}
According  to  Eq. \eqref{eq:2D_coin_embedded}, the $z$-rotation in $D=2$, when embedded in a coin space of dimension $D>2$, reads
\begin{equation}
\label{rezD}
R^{(E,D>2)}_{z}(\theta) = 
\operatorname{diag}\left( \{e^{- i\theta/2},1,\ldots,1,e^{+ i \theta/2}\}\right).
\end{equation}
The initial state of the system \eqref{eq:probe_state}, with initial coin state as in Eq. \eqref{eq:arbitrary_coin_state}, evolves according to
\begin{align}
    \vert \psi(t) \rangle =& \sum_{\sigma = \pm 1} e^{ i \sigma t \theta /2} \chi_{\sigma M} \vert \sigma t M \rangle_p \otimes  \vert \sigma M \rangle_c \nonumber\\
    &+ \sum_{-M < m < M} \chi_{m} \vert m t\rangle_p \otimes  \vert m \rangle_c,
    \label{eq:psit_REzD}
\end{align}
with $M$ in Eq. \eqref{eq:index_M_Rz}. The corresponding QFI \eqref{quantum_fisher_information},
\begin{align}
    H^{(E,D>2)}_{z}(t) = t^2 \Big[& \vert \chi_{-M} \vert^2 +\vert \chi_{+M}\vert^2 \nonumber\\
    &- \left(\vert \chi_{-M} \vert^2 - \vert \chi_{+M}\vert^2 \right)^2\Big] ,
    \label{eq:QFI_arbitrary_D_REzD}
\end{align}
is independent of $\theta$ and of the phases $\{\gamma_i\}_i$ of the probe state. Again, the FI is identically null because the reduced density matrix is independent of $\theta$ and so is the probability distribution of walker's position. 

\subsection{Explicit results for dimension $D=2$}
The spin-$1/2$ rotation ($D=2$) around the $z$-axis has matrix representation
\begin{equation}
R^{(2)}_{z}(\theta)=\begin{pmatrix}
e^{-i\theta/2} & 0 \\
0 & e^{i\theta/2}
\end{pmatrix},
\label{eq:Rz2_matrix}
\end{equation}
which leads to the QFI
\begin{equation}
\label{Rz2_qfi}
H^{(2)}_{z}(t) = t^2 \sin^{2} \alpha_{1} .
\end{equation}
It is minimum, $H^{(2)}_{z}=0$, for $\alpha_{1}=0,\pi$, i.e., when the probe is a coin basis state, and it is maximum, $H^{(2)}_{z}=t^2$, for $\alpha_{1}=\pi/2$ regardless of the phase $\gamma_{1}$, i.e., when the probe is optimal $\vert \Phi^{(2)}_{z} \rangle_c = \left(  \vert -1 \rangle_c + e^{i\gamma_{1}} \vert +1 \rangle_c \right)/\sqrt{2}$

As already proved, the FI of a position measurement is identically zero, independently of the initial state. From Eq. \eqref{eq:redrho_p_RzD}, we observe that
\begin{equation}
    \rho_p = \cos^2 \left(\frac{\alpha_1}{2}\right) \vert - t \rangle_p{}_p\langle - t \vert + \sin^2\left(\frac{\alpha_1}{2}\right) \vert + t \rangle_p{}_p\langle + t \vert, 
\end{equation}
which means that, at time $t$, the walker populates only the sites $x = \pm t$ with non-zero probability and we have the certainty of finding it in the site $x = -t$ ($x=+t$) if $\alpha_1 = 0$  ($\alpha_1 = \pi$).

\subsection{Explicit results for dimension $D=3$}
We compare the embedding of a two-dimensional $z$-rotation in a higher dimensional space, $D=3$, to the  actual three-dimensional $z$-rotation.
We embed $R_z^{(2)}(\theta)$ \eqref{eq:Rz2_matrix} into a $(D=3)$-dimensional space as
\begin{equation}
\label{erz3}
R_{z}^{(E,3)}(\theta)=\begin{pmatrix}
e^{-i\frac{\theta}{2}} & 0 & 0 \\
0 & 1 & 0 \\
0 & 0 & e^{+i\frac{\theta}{2}}
\end{pmatrix},
\end{equation}
which leads to the QFI
\begin{align}
\label{erz3_qfi}
H_{z}^{(E,3)}(t) & = t^2 \Bigl[ 2 \sin^{2}\left( \frac{\alpha_{1}}{2} \right)\cos^{2}\left( \frac{\alpha_{2}}{2} \right)\cos^{2}\left( \frac{\alpha_{1}}{2} \right)  \nonumber \\
&\quad + \sin^{2}\left( \frac{\alpha_{1}}{2} \right)\cos^{2}\left( \frac{\alpha_{2}}{2} \right) + \cos^{2}\left( \frac{\alpha_{1}}{2} \right) \nonumber \\
&\quad  -\sin^{4}\left( \frac{\alpha_{1}}{2} \right)\cos^{4}\left( \frac{\alpha_{2}}{2} \right) - \cos^{4}\left( \frac{\alpha_{1}}{2} \right)\Bigr].
\end{align}

The maximum QFI 
\begin{equation}
    \max_{\vert \phi^{(2)} \rangle_c} H^{(2)}_{z}(t) = \max_{\vert \phi^{(3)} \rangle_c} H_{z}^{(E,3)}(t) = t^2
    \label{eq:maxQFI_erz3_rz2}
\end{equation}
is achieved for the optimal probe with $\alpha_1 = \pi/2$ and $\alpha_2 = 0$, $\vert \Phi_z^{(E,3)} \rangle_{c} =  ( \vert -1 \rangle_c  + e^{i\gamma_{2}} \vert +1 \rangle_c )/\sqrt{2}$. This result, together with Eq. \eqref{eq:QFI_arbitrary_D_REzD}, reveals that embedding a $R_z^{(2)}(\theta)$ coin rotation in a higher dimensional space does not improve the maximum achievable QFI. Even in this case, the FI for walker's position measurement is null. Therefore, a higher dimensional coin is not a metrological resource when embedded coin operators are considered.

On the other hand, a higher dimensional coin space is a resource for simulating DTQWs generated by a lower dimensional coin. E.g., the DTQW generated by the embedded coin $R_z^{(E,3)}(\theta)$ \eqref{erz3} with initial coin state \eqref{3D_Bloch_Sphere} with $\alpha_2 = 0$ is equivalent to the DTQW generated by the coin $R_z^{(2)}(\theta)$ \eqref{eq:Rz2_matrix} with initial coin state \eqref{Bloch_Sphere}. Accordingly, they also provide the same QFI (see Eq. \eqref{erz3_qfi}, with $\alpha_2=0$, and Eq. \eqref{Rz2_qfi}).

The actual spin-$1$ rotation ($D=3$) around the $z$-axis has matrix representation
\begin{equation}
\label{R3}
R^{(3)}_{z}(\theta)=\begin{pmatrix}
e^{-i\theta} & 0 & 0 \\
0 & 1 & 0 \\
0 & 0 & e^{i\theta}
\end{pmatrix}.
\end{equation}
It differs from the embedded rotation \eqref{erz3} only by a factor two in the argument of the exponential. Therefore, the resulting QFI is $H^{(3)}_{z}(t) = 4 H_{z}^{(E,3)}(t)$, (see Eq. \eqref{erz3_qfi}), its maximum value is $\max_{} H^{(3)}_{z}(t) = 4t^2$, (see Eq. \eqref{eq:maxQFI_erz3_rz2} and Eq. \eqref{max_QFI_Rz}), and the optimal probe is  the same $\vert \Phi^{(3)}_{z} \rangle_c = \vert \Phi^{(E,3)}_{z} \rangle_c$.

\subsection{Explicit results for dimension $D=4$} 
We compare the embedding of a two-dimensional $z$-rotation in a higher dimensional space, $D=4$, to the actual four-dimensional $z$-rotation.
We embed $R_z^{(2)}(\theta)$ \eqref{eq:Rz2_matrix} into a $(D=4)$-dimensional space as
\begin{equation}
\label{erz4}
R_{z}^{(E,4)}(\theta)  = 
\begin{pmatrix}
e^{-i\frac{\theta}{2}} & 0 & 0 & 0 \\
0 & 1 & 0 & 0 \\
0 & 0 & 1 & 0 \\
0 & 0 & 0 & e^{i\frac{\theta}{2}} 
\end{pmatrix}.
\end{equation}
The resulting QFI is $H_{z}^{(E,4)}(t) = H_{z}^{(E,3)}(t)$, (see Eq. \eqref{erz3_qfi}). This equality follows from the parametrization of the coin state in Eq. \eqref{D_Bloch_Sphere} and the embedding defined in Eq. \eqref{eq:2D_coin_embedded} for arbitrary $D$, as the result involves the same variables.

The actual spin-$3/2$ rotation around the $z$-axis has matrix representation
\begin{equation}
\label{R4}
R^{(4)}_{z}(\theta) = 
\begin{pmatrix}
e^{-i\frac{3\theta}{2}} & 0 & 0 & 0 \\
0 & e^{-i\frac{\theta}{2}} & 0 & 0 \\
0 & 0 & e^{i\frac{\theta}{2}} & 0 \\
0 & 0 & 0 & e^{i\frac{3\theta}{2}} 
\end{pmatrix},
\end{equation}
which leads to the QFI

\begin{align}
H^{(4)}_{z}(t)  = & t^2 \biggl\{ 9 \left[ \sin^{2}{\left( \frac{\alpha_{1}}{2} \right)}\cos^{2}{\left( \frac{\alpha_{2}}{2} \right) } + \cos^{2}{\left( \frac{\alpha_{1}}{2} \right)} \right] \nonumber \\
& - \left[ 3 \left( \sin{\left( \frac{\alpha_{1}}{2} \right)}^{2}\cos{\left( \frac{\alpha_{2}}{2} \right)}^{2} - \cos^{2}{\left( \frac{\alpha_{1}}{2} \right)} \right) \right. \nonumber \\  
& \left. + \sin^{2}{\left( \frac{\alpha_{1}}{2} \right)}\sin^{2}{\left( \frac{\alpha_{2}}{2} \right)}\left(1-2\sin^{2}{\left( \frac{\alpha_{3}}{2} \right)} \right)\right] ^{2} \nonumber \\
& + \sin^{2}{\left( \frac{\alpha_{1}}{2} \right)}\sin^{2}{\left( \frac{\alpha_{2}}{2} \right) } \biggr\} 
\end{align}
whose maximum, $H^{(4)}_{z}=9t^2$, is achieved for $\alpha_{1}=\pi / 2$ and $\alpha_{2}=0$, irrespective of $\alpha_{3}$ and of $\{ \gamma_i \}_i$, i.e., when the probe is optimal $\vert \Phi^{(4)}_{z} \rangle_c = \left(  \vert -2 \rangle_c + e^{i\gamma_{3}} \vert +2 \rangle_c \right)/\sqrt{2}$.

\section{Metrology with $x$- and $y$-rotations encoding and different coin dimensions}
\label{s:xy}
The $x$- and $y$-rotation operators are not diagonal in the chosen basis for the coin space. Therefore, we numerically study the corresponding DTQW and the associated estimation problem. We will further inspect these results for the specific values of $\theta$ for which we can provide analytical results.
As discussed in Sec. \ref{Main_Results}, the embedding of a two-dimensional $z$-rotation in a higher dimensional space turns out not to be of metrological interest, as it does not improve the QFI. Therefore, in the following we focus exclusively on the actual $x$- and $y$-rotations. Before discussing \textcolor{black}{ in detail our} results for dimensions $D= 2, 3, 4$, we list here the features of the QFI that we found to be common to $R_{x}^{(D)}$ and $R_{y}^{(D)}$ and that cut across all the dimensions $D=2,3,4$: (i) The QFI does depend on $\theta$, unlike for $z$-rotations; for small angles, $\theta \to 0$, (ii) the QFI is linear in time, while for finite angles the leading term is quadratic in time,
\begin{equation}
    \lim_{\theta \to 0} H_{x,y}^{(D)}(\theta) \propto t, \quad \text{vs} \quad  H_{x,y}^{(D)}(\theta\gtrsim 0) \propto t^2,
\end{equation}
and (iii) the FI approaches the QFI,
\begin{equation}
    \lim_{\theta \to 0}\, \max_{\vert \phi^{(D)} \rangle_c} F_{X; x,y}^{(D)} (\theta) = \lim_{\theta \to 0}\, \max_{\vert \phi^{(D)} \rangle_c} H^{(D)}_{x,y}(\theta),
\end{equation}
meaning that walker's position measurement is nearly optimal for estimating $\theta$. As a downside, the latter result also means that such a measurement can extract the maximum information available on $\theta$ only when the QFI is low compared to that for other values of $\theta$. In addition, we have that
\begin{equation}
    \max_{\vert \phi^{(D)} \rangle_c} H_x^{(D)}(t,\theta) = \max_{\vert \phi^{(D)} \rangle_c} H_y^{(D)}(t,\theta).
\end{equation}
In principle, the QFI for $x$-rotations and that for $y$-rotations are maximized by different optimal probes. However, it is possible to find states that are simultaneously optimal for both the rotations.

\subsection{Explicit results for dimension $D=2$}
The spin-$1/2$ rotation ($D=2$) around the $x$- or $y$-axis has matrix representation
\begin{equation}
\label{eq:Rxy_D2}
    R_n^{(2)}(\theta) =  \cos(\theta/2) \mathbb{1}  -i 2 \sin(\theta/2) \mathcal{T}_n^{(2)},
\end{equation}
where the generators $\mathcal{T}_{n}^{(2)}$, with $n=x,y$, are defined in  Eq. \eqref{tx_2D}--\eqref{ty_2D}. In the following, first we consider the $y$-rotations, then the $x$-rotations.

\begin{figure}[h!]
    \includegraphics[width=\columnwidth]{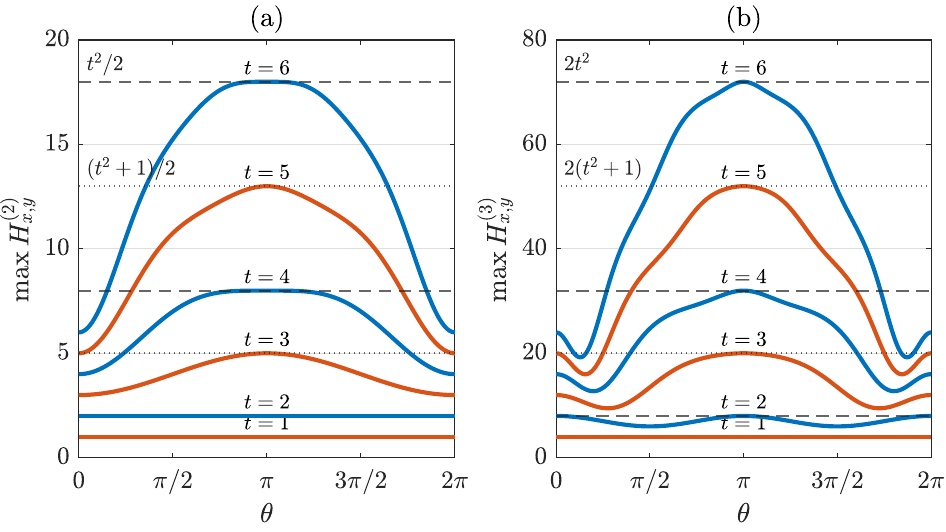}
    \caption{Maximum QFI as a function of $\theta$ for rotations $R_{x,y}^{(D)}(\theta)$, for time steps $t=1,2,\ldots 6$. (a) Results for $D=2$ and optimal probe $\vert \Phi_{x,y}^{(2)} \rangle_c = \vert -1 \rangle_c$ ($\alpha_1 = 0$). For $\theta \to 0$ the QFI is $H_{x,y}^{(2)}(0) = t$ [Eq. \eqref{eq:suy2_qfi_0}]. For $\theta = \pi$ the QFI takes its maximum value $H_{x,y}^{(2)}(\pi)= [t^2 + \bmod(t,2)]/2$, that differs between odd times [orange line, Eq. \eqref{suy2_theta=pi/2_qfi_odd}] and even times [blue line, Eq. \eqref{suy2_theta=pi/2_qfi_even}]. (b) Same quantities for $D=3$ and optimal probe $\vert \Phi_{x,y}^{(3)} \rangle_c = \vert 0 \rangle_c$ ($\alpha_1 = \alpha_2 = \pi$). The QFI is $H_{x,y}^{(3)}(\theta) = 4H_{x,y}^{(2)}(\theta)$ for $\theta = 0,\pi$. The QFI shows a period of $2 \pi$.}
    \label{fig:Rxy_max_qfi_23D}
\end{figure}

\textit{(i) $y$-rotations.---}We can analytically inspect the QFI for some peculiar angles to determine its exact expression. We start by considering the limit for $\theta \to 0$, i.e., a small deviation from the identity coin (see Eq. \eqref{eq:Rxy_D2}). In this regime, the QFI is linear in $t$,
\begin{equation}
\label{eq:suy2_qfi_0}
\lim_{\theta \to 0}H^{(2)}_{y} (\theta,t) = t - \sin^{2}\alpha_{1}\sin^{2}\gamma_{1},
\end{equation}
while, for large $t$, it recovers the quadratic behavior for $\theta \gtrsim 0$ (see Fig. \ref{fig:Rxy_max_qfi_23D}(a) for $\theta \approx 0$ and $\theta \approx \pi$).  The optimal probe that maximizes both the FI and the QFI is the state \eqref{Bloch_Sphere} with $\gamma_1 = 0,\pi$ or $\alpha_1 = 0,\pi$ (see Eq. \eqref{eq:suy2_qfi_0}). Similar results are found for $\theta = 2\pi$, as the rotation is $R_y^{(2)}(2\pi) = -R_y^{(2)}(0)$ (see Eq. \eqref{eq:Rxy_D2}), so the difference with respect to $\theta = 0$ is just a phase $(-1)^t$ in the state, $\vert \psi_{\theta = 2 \pi}(t) \rangle  = (-1)^t \vert \psi_{\theta = 0}(t) \rangle$, and thus in its derivative. Such phases compensate when computing the QFI \eqref{quantum_fisher_information}.

Another peculiar value is $\theta = \pi$ (or $\theta = 3\pi$), for which the rotation is $R_y^{(2)}(\pi) = -i\sigma_y$, with $\sigma_y$ the Pauli matrix. Accordingly, the system oscillates between two configurations: localized in $\vert 0 \rangle_p$ for even $t$, with corresponding QFI
\begin{align}
\label{suy2_theta=pi/2_qfi_even}
H^{(2)}_{y} (\text{even } t) = \frac{t^2}{2} \left[ 1-\frac{1}{2}\sin^{2}\gamma_{1} \sin^{2} \alpha_{1} \right],
\end{align}
and delocalized in its two nearest neighbours $\vert \pm 1 \rangle_p$ for odd $t$, with corresponding QFI
\begin{align}
\label{suy2_theta=pi/2_qfi_odd}
H^{(2)}_{y} (\text{odd } t) = \left[ \frac{t^2+1}{2} - \frac{(t+1)^2}{4}\sin^{2} \gamma_{1} \sin^{2} \alpha_1  \right].
\end{align}
For $\theta = 3\pi$, the rotation is $R_y^{(2)}(3\pi) = i\sigma_y$, so the difference with respect to $\theta = \pi$ is just a phase $(-1)^t$ in the state and its derivative, which compensate when computing the QFI \eqref{quantum_fisher_information}.

For $R^{(2)}_{y}(\theta)$, any real state,
\begin{align}
\vert \Phi^{(2)}_{y} \rangle_c = \cos{\left( \frac{\alpha_{1}}{2} \right)} \vert -1 \rangle_c \pm \sin{\left( \frac{\alpha_{1}}{2} \right)} \vert +1 \rangle_c,
\label{state_max_y_rot_x_y_2D}
\end{align}
is optimal and leads to the same maximum QFI. Being both $R_y^{(2)}(\theta)$ and the probe real, the maximum QFI simplifies to (see Appendix \ref{app:psi_dpsi_orthogonality})
\begin{equation}
\label{quantum_fisher_information_real}
H_y^{(2)}(\theta) = 4 \langle \partial_{\theta} \psi_{\theta} \vert \partial_{\theta} \psi_{\theta} \rangle .
\end{equation}
For $y$-rotation in $D=2$, although the QFI depends on $\theta$ [see, e.g., Fig. \ref{fig:Rxy_max_qfi_23D}(a)], its maximization over the possible probes is independent of $\theta$ and the maximum QFI is independent of the initial state, provided it is real. So, as for $R^{(2)}_{z}(\theta)$, there is a family of optimal states for the QFI. However, if for $z$-rotation the condition was to have the angle $\alpha_{1}$ fixed and the phase $\gamma_{1}$ free, for $y$-rotation it is to have the phase $\gamma_{1}$ fixed and the angle $\alpha_{1}$ free.

\textit{(ii) $x$-rotations.---}The exact QFI for the peculiar angles $\theta = 0,2\pi$ and $\theta = \pi,3\pi$ can be derived from that obtained for the $y$-rotation upon replacing $\sin^2{\gamma_{1}} \to \cos^2{\gamma_{1}}$, in Eq. \eqref{eq:suy2_qfi_0} and Eqs. \eqref{suy2_theta=pi/2_qfi_even}--\eqref{suy2_theta=pi/2_qfi_odd}, respectively. The optimal probes that maximize the QFI are the states \eqref{Bloch_Sphere} with $\gamma_1 = \pi/2, 3\pi/2$ or $\alpha_1 = 0,\pi$, conditions that can be summarized by the state
\begin{align}
\label{state_max_x_rot_x_y_2D}
&\vert \Phi^{(2)}_{x} \rangle_c = \cos{\left( \frac{\alpha_{1}}{2} \right)} \vert -1 \rangle_c \pm i \sin{\left( \frac{\alpha_{1}}{2} \right)} \vert +1 \rangle_c.
\end{align}
We have therefore determined the probes that separately maximize all the three rotations $R_n^{(2)}(\theta)$, with $n=x,y,z$. It is not possible to simultaneously maximize the QFI of these three rotations as they require incompatible conditions on the optimal probe state: $\alpha_1 = \pi/2$ for the $z$-rotation,  $\alpha_1 = 0,\pi$ or $\gamma_1 = 0$ for the $y$-rotation, and $\alpha_1 = 0,\pi$ or $\gamma_1 = \pi/2,3\pi/2$ for the $x$-rotation. However, it is possible to simultaneously maximize the QFI for two rotations with the following probes
\begin{align}
&\vert \Phi^{(2)}_{x,z} \rangle_c = \frac{1}{\sqrt{2}} \left(  \vert -1 \rangle_c \pm i \vert +1 \rangle_c \right), \label{initial_max_QFI_Rz2_Rx2} \\
&\vert \Phi^{(2)}_{x,y} \rangle_c = \vert \pm 1 \rangle_c, \label{initial_max_QFI_Rx2_Ry2}\\
&\vert \Phi^{(2)}_{y,z} \rangle_c = \frac{1}{\sqrt{2}} \left(  \vert -1 \rangle_c \pm \vert +1 \rangle_c \right) \label{initial_max_QFI_Rz2_Ry2},
\end{align}
where the subscripts denote the two rotations whose QFI is maximized. As a final remark, we point out that the QFI for $R^{(2)}_{x}(\theta)$ or $R^{(2)}_{y}(\theta)$ is always lower than that for $R^{(2)}_{z}(\theta)$, see, e.g., Fig. \ref{fig:max_qfi_Rxyz}. On the other hand, the FI for $x$- and $y$-rotations is nonzero, thus higher than that for $z$-rotation (for which it vanishes, \textcolor{black}{due to the structure of Eq. \eqref{eq:redrho_p_RzD}). The FI of a $z$-rotation is identically null because the probability distribution for each initial state does not depend on $\theta$ and then, any position measurement can not infer information about the coin parameter. On the contrary, for $x$- and $y$-rotations, the interference phenomena in position space results in a reduction in the overall Quantum Fisher Information of the system. Despite this reduction, the probability distribution in the walker's space is then dependent on $\theta$. Consequently, measuring the walker's position can indeed infer information about $\theta$. Therefore, at the cost of a diminished global QFI, there exists a non-zero position FI.}

\begin{figure}[h!]
    \includegraphics[width=\columnwidth]{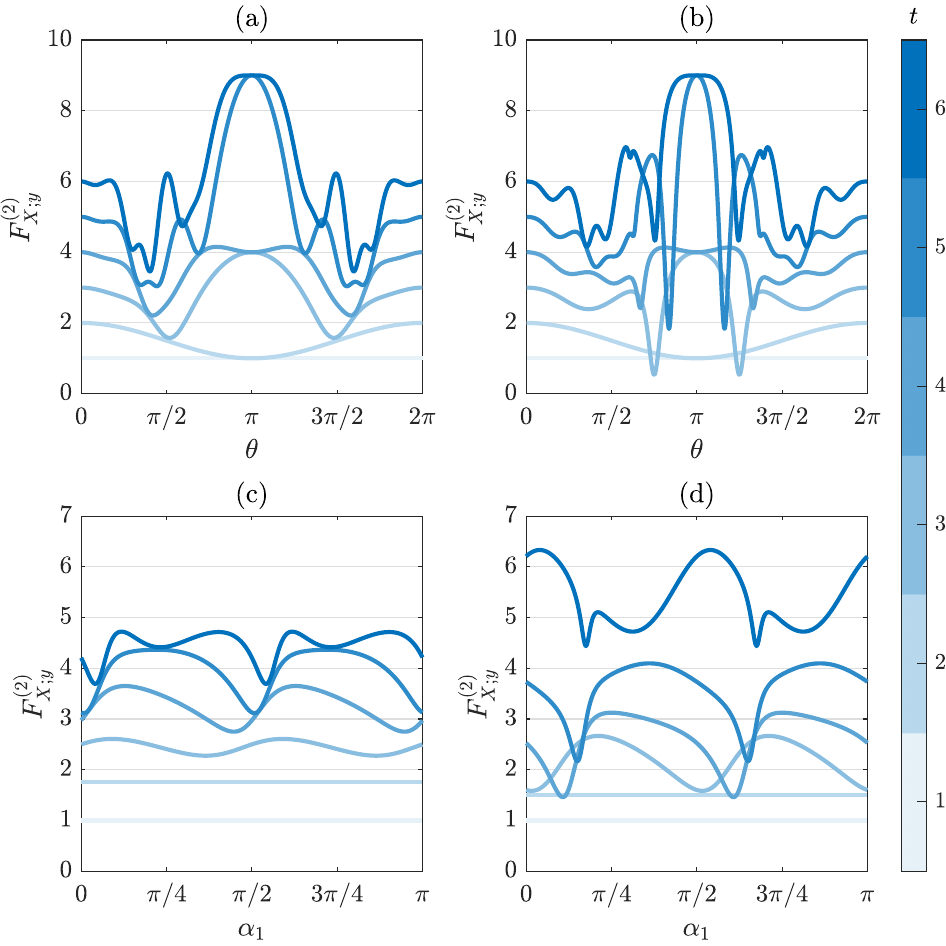}
    \caption{Classical FI of a position measurement for rotation $R_{y}^{(2)}(\theta)$ for time steps $t=1,2,\ldots,6$. Panels (a,b): Results as a function of $\theta$ for the initial state \eqref{Bloch_Sphere} with $\gamma_1 = 0$ and (a) $\alpha_{1}=0$ and (b) $\alpha_{1}=\pi/4$ [both optimal states, Eq. \eqref{state_max_y_rot_x_y_2D}]. Panels (c,d): Results for (c) $\theta=\pi/3$ and (d) $\theta=\pi/2$ as a function of the angle $\alpha_1$ parametrizing the optimal, real initial state \eqref{state_max_y_rot_x_y_2D} with the plus sign [i.e., $\gamma_1=0$ in Eq. \eqref{Bloch_Sphere}]. In all panels, $F_{X;y}^{(2)}(t=1)\equiv 1$.}
    \label{fig:FI_Ry_2D}
\end{figure}

Let us now focus on the FI for the optimal states maximizing the QFI. In the limit of $\theta \to 0$, the FI equals the QFI, is linear in time [see, e.g., Fig. \ref{fig:Rxy_max_qfi_23D}(a) and \ref{fig:FI_Ry_2D}(a)], and thus the optimal states for the QFI also maximize the FI. Unlike the QFI, the FI does not show analogous regularities in the time dependence and in the state maximizing the FI. As the time increases, the maxima of the FI do not occur always at the same value of $\theta$ and, at given $\theta$, the FI is not necessarily increasing in time, i.e., it is possible to have $F_{X}(\theta^\ast,t)>F_{X}(\theta^\ast,t+1)$ [see Fig. \ref{fig:FI_Ry_2D}(a,b)]. This behavior is in sharp contrast with that of the QFI, whose maximum occurs at $\theta=\pi$ and, at given $\theta$, it is increasing in time [Fig. \ref{fig:Rxy_max_qfi_23D}(a)]. In addition, the FI strongly depends on the initial real state \eqref{state_max_x_rot_x_y_2D} considered [Fig. \ref{fig:FI_Ry_2D}(c,d)], while the value of the QFI is maximum and independent of it.

\subsection{Explicit results for dimension $D=3$}
The spin-$1$ rotation ($D=3$) around the $x$- or $y$-axis has matrix representation
\begin{equation}
\label{eq:Rxy_D3}
 R_n^{(3)}(\theta) =  \mathbb{1} -i \sin(\theta/2) \mathcal{T}_n^{(3)} + (\cos(\theta/2) -1) {\mathcal{T}_n^{(3)}}^2,
\end{equation}
where the generators $\mathcal{T}_{n}^{(3)}$, with $n=x,y$, are defined in  Eq. \eqref{tx_3D}--\eqref{ty_3D}. Again, we can analytically study the QFI for specific angles to inspect the metrological advantage---higher QFI---with respect to the case $D=2$. In particular, we observe that
\begin{align}
    \lim_{\theta \to 0}\left[ \max_{\vert \phi^{(3)} \rangle_c} H_{x,y}^{(3)} \right] &= 4 \lim_{\theta \to 0} \left[ \max_{\vert \phi^{(2)} \rangle_c} H_{x,y}^{(2)} \right] = 4t,\\
    \max_{\theta}\left[ \max_{\vert \phi^{(3)} \rangle_c} H_{x,y}^{(3)} \right] &= 4 \max_{\theta} \left[ \max_{\vert \phi^{(2)} \rangle_c} H_{x,y}^{(2)} \right] \nonumber\\
    &= 2[t^2+\bmod(t,2)].
\end{align}
This means that the QFI corresponding to the optimal probe, intended as the probe which provides the highest $\max_\theta H_{x,y}^{(3)}$, is enhanced by a factor $4$ compared to the case in $D=2$, the same improvement we had for the $z$-rotation when passing from $D=2$ to $D=3$, (see Eq. \eqref{max_QFI_Rz}).  However, for $x$- and $y$-rotations this gain does not hold for all the values of $\theta$, but for $\theta \to 0$ and $\theta = \pi$ [compare Fig. \ref{fig:Rxy_max_qfi_23D}(a,b)]. The optimal probe is unique and simultaneously optimizes the QFI for both the rotations 
\begin{align}
\label{initial_max_QFI_Rxy3D}
\vert \Phi^{(3)}_{x,y} \rangle_c = \vert 0 \rangle_c.
\end{align}
Here is a major difference with respect to the case in $D=2$: \textcolor{black}{The optimal probe (i) is unique and (ii) is simultaneously optimal only for x- and y-rotations. Such a probe does not optimize the QFI for z-rotations because the optimal probe of the latter is Eq. \eqref{initial_max_QFI_Rz} with M = 1.} It is worth noticing that in the $y$ case even if the coin matrix is real, not \textcolor{black}{all} real states maximize the QFI. Even if the QFI reduces to Eq. \eqref{quantum_fisher_information_real}, in $D=3$ the square modulus of the derivative of the wave function is not constant for any real state. This means that the orthogonality of a state and its derivative, $\langle \partial_\theta \psi_\theta \vert \psi_\theta \rangle=0$, is not a sufficient condition for the maximization of the QFI (see Eq. \eqref{quantum_fisher_information}).

\begin{figure}[h!]
    \includegraphics[width=\columnwidth]{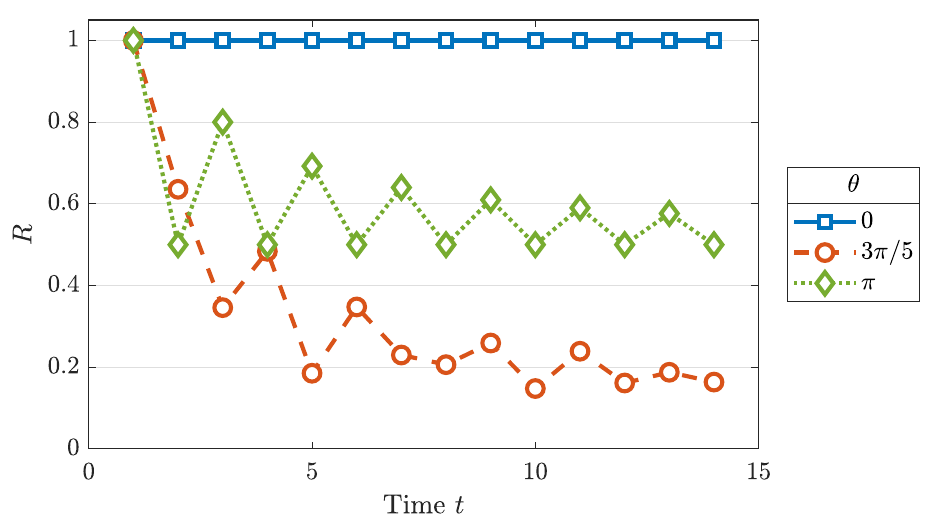}
    \caption{Ratio $R$ between the FI and the QFI  as a function of the time step for the optimal probe $\vert \Phi^{(3)}_{x,y} \rangle_c = \vert 0 \rangle_c$ \eqref{initial_max_QFI_Rxy3D} and different values of $\theta$.}
    \label{fig:R_roty_D=3}
\end{figure}

The ratio $R=F_X/H$ between FI and QFI is indicative of the optimality of a measurement $X$. Indeed, it is bounded by $0 \leq R \leq 1$, with $R=1$ for optimal measurements (see Eq. \eqref{eq:QFI_def_geq_FI}). Focusing on the FI of a position measurement performed on the optimal state \eqref{initial_max_QFI_Rxy3D} for the QFI, we observe in Fig. \ref{fig:R_roty_D=3} that the ratio strongly depends on the value of $\theta$ and on the time step (except for $\theta=0$). Therefore, the amount of information encoded on the outcomes of a position measurement depends on $\theta$ and $t$. \textcolor{black}{The suitability of a position measurement to estimate the parameter of interest depends on both $\theta$ and $t$ and it is quantified by the values of $R(\theta,t)$: A position measurement is nearly optimal (poor) for the value of $\theta$ and $t$ for which $R \approx 1$ ($R \approx 0$).} Numerical analysis suggests the existence of asymptotic value $\lim_{t \to \infty}R = R_\theta$ (achieved from below for even $t$ and from above for odd $t$), which however strongly depends on $\theta$.

\subsection{Explicit results for dimension $D=4$}
For the spin-$3/2$ rotation ($D=4$) around the $x$- or $y$-axis, we observe that
\begin{align}
    \lim_{\theta \to 0}\left[ \max_{\vert \phi^{(4)} \rangle_c} H_{x,y}^{(4)} \right] &= 7 \lim_{\theta \to 0} \left[ \max_{\vert \phi^{(2)} \rangle_c} H_{x,y}^{(2)} \right] = 7t,\\
    \max_{\theta}\left[ \max_{\vert \phi^{(4)} \rangle_c} H_{x,y}^{(4)} \right] &= 7\max_{\theta} \left[ \max_{\vert \phi^{(2)} \rangle_c} H_{x,y}^{(2)} \right] \nonumber\\
    &= \frac{7}{2}[t^2+\bmod(t,2)].
\end{align}
The highest $\max_\theta H_{x,y}^{(3)}$ is enhanced by a factor $7$ compared to the case in $D=2$, unlike the improvement by a factor $9$ we had for the $z$-rotation when passing from $D=2$ to $D=4$, (see Eq. \eqref{max_QFI_Rz}).

As in the case $D=2$ we have more than one optimal state for $R^{(4)}_{y}(\theta)$ and $R^{(4)}_{x}(\theta)$, that respectively read
\begin{align}
&\vert \Phi^{(4)}_{y} \rangle_c = \cos{\left( \frac{\alpha_{3}}{2} \right)} \vert -1 \rangle_c \pm \sin{\left( \frac{\alpha_{3}}{2} \right)} \vert +1 \rangle_c, \label{state_max_x_rot_y_4D}\\
&\vert \Phi^{(4)}_{x} \rangle_c = \cos{\left( \frac{\alpha_{3}}{2} \right)} \vert -1 \rangle_c \pm i \sin{\left( \frac{\alpha_{3}}{2} \right)} \vert +1 \rangle_c. \label{state_max_x_rot_x_4D}
\end{align}
Again, by optimal probe we mean the probe which provides the highest $\max_\theta H_{x,y}^{(4)}$. The optimal probes resemble the optimal ones we determined for $R^{(2)}_{y}(\theta)$ and $R^{(2)}_{x}(\theta)$, Eq. \eqref{state_max_y_rot_x_y_2D} and \eqref{state_max_x_rot_x_y_2D}, respectively. Analogously to the case $D=3$, it is not possible to simultaneously maximize the QFI for both $x(y)$- and $z$-rotations, but the state
\begin{equation}
\label{initial_max_QFI_Rx4_Ry4}
\vert \Phi^{(4)}_{x,y} \rangle_c = \vert \pm 1 \rangle_c
\end{equation}
maximizes both the $x$- and $y$- rotation, similarly to the case ${D}=2$ (see Eq. \eqref{initial_max_QFI_Rx2_Ry2}).
As for the $R_{x,y}^{(2)}(\theta)$ when we have a family of states that maximize $H_{x,y}^{(D)}(\theta)$ we cannot find optimal state(s) for $F_{X}$ valid for each time step $t$ or for each value of $\theta$.

\begin{figure}[h!]
    \includegraphics[width=\columnwidth]{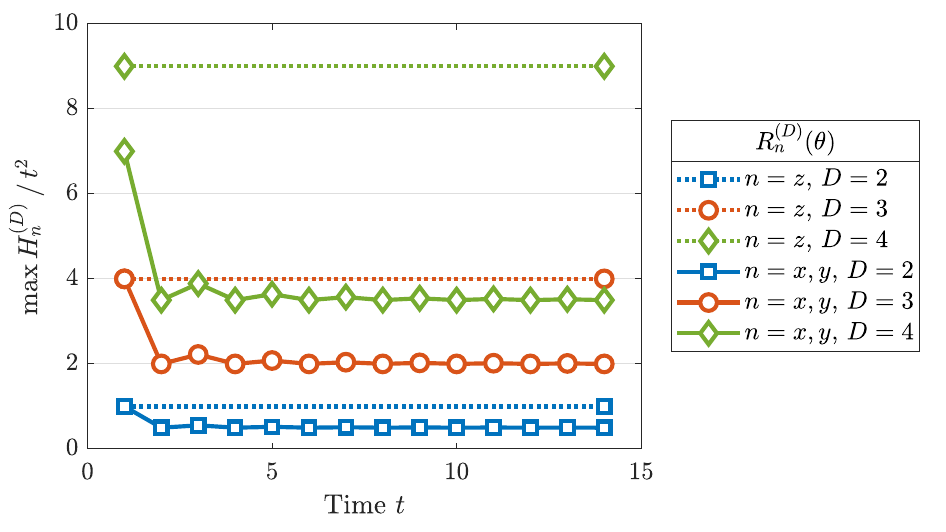}
    \caption{Asymptotic behavior of the QFI maximized over the initial states \eqref{eq:probe_state} and values of $\theta$ for rotations $R_{n}^{(D)}(\theta)$, with $n=x,y,z$ and $D=2,3,4$. For large enough time, the maximum QFI is $\sim O(t^2)$.}
    \label{fig:max_qfi_Rxyz}
\end{figure}

To conclude the discussion on the estimation problem for rotation encodings, we compare the asymptotic behavior of the maximum QFI, optimized over the probe states and $\theta$, for the rotations $R_n^{(D)}(\theta)$
\begin{equation}
    \lim_{t\to\infty} \max_{\vert \phi^{(D)} \rangle_c} \frac{H_n^{(D)}}{t^2} =
    \begin{cases}
    (D-1)^2 & \text{for $n=z$,}\\
    1/2 & \text{for $n=x,y$ and $D=2$,}\\
    2 & \text{for $n=x,y$ and $D=3$,}\\
    7/2 & \text{for $n=x,y$ and $D=4$.}
    \end{cases}
\end{equation}
This result reveals that, at least in $D=2,3,4$, the QFI for $z$-rotations is always larger than that for $x,y$-rotations.
This long-time limit regime is achieved after a few time steps (Fig. \ref{fig:max_qfi_Rxyz}).

\section{Metrology with generalized Grover walk encoding and different coin dimensions}
\label{s:gr}
In this section we discuss the estimation problem of the single parameter of the generalized Grover coin for $D=2,3$, \textcolor{black}{ defined in Sec. \ref{sec:coin_ops}}. The Grover coin in $D=2$ turns out to be of interest also because it can be read as the composition of two rotations previously considered.

\begin{figure}[h!]
    \includegraphics[width=\columnwidth]{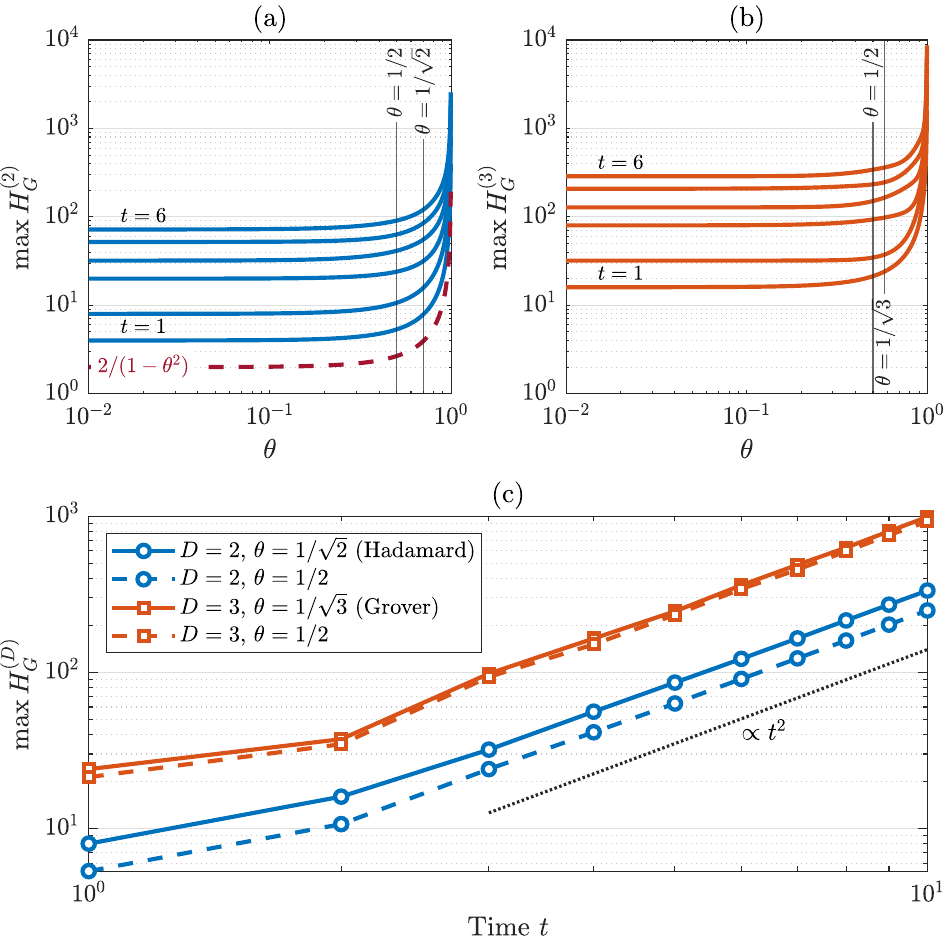}
    \caption{Maximum QFI in a generalized Grover walk for $D=2,3$. (a) QFI in $D=2$ as a function of $\theta$ for $t=1,\ldots,6$ (blue lines). The curve $2/(1-\theta^2)$ (red dashed line) is reported as a reference. (b) QFI in $D=3$ as a function of $\theta$ for $t=1,\ldots,6$ (orange lines). (c) QFI for the Hadamard walk ($D=2$, $\theta = 1/\sqrt{2}$), Grover walk ($D=3$, $\theta = 1/\sqrt{3}$), and comparison of the QFI for $D=2,3$ at fixed $\theta = 1/2$. The asymptotic behavior of the QFI is $O(t^2)$. Results obtained for the optimal probes $\vert \Phi_G^{(2)} \rangle_c = \vert -1 \rangle_c$ and $\vert \Phi_G^{(3)} \rangle_c = \vert 0 \rangle_c$.}
    \label{fig:G_max_qfi}
\end{figure}

\textit{(i) Dimension $D=2$.---} The QFI associated to the generalized Grover walk with coin \eqref{2D_grover} depends on the parameter $\theta$ as 
\begin{equation}
\label{QFI_Grover_Dependence}
H_{G}^{(2)}(\theta,t) = \dfrac{f(\theta,t)}{1-\theta^2},
\end{equation}
which diverges for $\theta \to 1$, where $f(\theta,t)$ is polynomial in $\theta$ and $t$ [Fig. \ref{fig:G_max_qfi}(a)]. The probe \eqref{Bloch_Sphere} is optimal for $\gamma_{2} = 0,\pi$ or $ \alpha_{1} = 0,\pi$ (see Appendix \ref{app:psi_dpsi_orthogonality}), so it reads
\begin{align}
\label{state_max_grover_2D}
&\vert \Phi^{(2)}_{G} \rangle_c = \cos{\left( \frac{\alpha_{1}}{2} \right)} \vert -1 \rangle_c \pm \sin{\left( \frac{\alpha_{1}}{2} \right)} \vert +1 \rangle_c,
\end{align}
neglecting an overall phase factor. These optimal states provide the same QFI,  are optimal for any value of $\theta$, and are the same optimal states for $y$-rotation in $D=2$. In this regard, we recall that the Grover coin is the product of a $y$-rotation and a constant $z$-rotation. This can be easily verified by reparametrizing the coin \eqref{2D_grover} according to $\theta \in [0,1] \mapsto \cos {(} \tilde{\theta}/2 {)} \in [0,1]$ as follows
\begin{align}
\mathcal{C}_{G}^{(2)}(\tilde{\theta})
&=\begin{pmatrix}
\cos \big(\tilde{\theta}/2\big) & \sin\big(\tilde{\theta}/2\big) \\
\sin\big(\tilde{\theta}/2\big) & -\cos\big(\tilde{\theta}/2\big) \\
 \end{pmatrix} \nonumber\\
 &=\begin{pmatrix}
\cos\big(\tilde{\theta}/2\big) & -\sin\big(\tilde{\theta}/2\big) \\
\sin\big(\tilde{\theta}/2\big) & \cos\big(\tilde{\theta}/2\big) \\
 \end{pmatrix}
 \begin{pmatrix}
1 & 0 \\
0 & -1\\
 \end{pmatrix}
\nonumber\\
&=i R_y^{(2)}(\tilde{\theta}) R_z^{(2)}(\pi),
\label{eq:Grover_rot_2D}
\end{align}
where $0 \leq \tilde{\theta} \leq \pi$ to ensure that both sine and cosine are positive [see also Eqs. \eqref{eq:Rz2_matrix} and \eqref{eq:Rxy_D2}]. Estimating the parameter $0 \leq \theta \leq 1$ of the Grover coin amounts to estimating the value of $0 \leq \cos {(}\tilde{\theta}/2{)} \leq 1$ in Eq. \eqref{eq:Grover_rot_2D}.

\textit{(ii) Dimension $D=3$.---}
The optimal probe
\begin{equation}
\label{initial_max_QFI_Grover3D}
\vert \Phi^{(3)}_{G} \rangle_c = \vert 0 \rangle_c
\end{equation}
is unique and it also maximizes the QFI for $x$- and $y$-rotation in $D=3$, Eq. \eqref{initial_max_QFI_Rxy3D}. The QFI shows an analogous dependence on $\theta$ as in $D=2$, Eq. \eqref{QFI_Grover_Dependence}, showing the same divergence [Fig. \ref{fig:G_max_qfi}(b)]. 

Results suggest that the QFI is basically independent of $\theta$ when the latter is small and that it is of order $O(t^2)$ for large enough time. For a given value of $\theta$, the maximum value the QFI can reach is higher in $D=3$ than in the $D=2$ case [Fig. \ref{fig:G_max_qfi}(a,b)]. Panel (c) shows in detail this comparison for $\theta = 1/2$, together with QFI for the Hadamard ($D=2$, $\theta = 1/\sqrt{2}$) and Grover walk ($D=3$, $\theta = 1/\sqrt{3}$).

\textcolor{black}{Furthermore, it is worth noticing that, in principle, different states of the bipartite system (walker’s position + coin) might result in the same probability distribution $p(x,\theta)$ of finding the walker in position x when the parameter takes the value $\theta$. Accordingly, for a given probability distribution $p(x,\theta)$ we may expect different values of the QFI, because the latter, by definition in Eq. \eqref{quantum_fisher_information}, depends on the quantum state and the derivative of the latter, not on $p(x,\theta)$.} Then, the probability distribution, or any physical quantity derived from $p(x|\theta)$, is not reliable to investigate the QFI when comparing different coins. As an example, if we take the ${D}=2$ generalized Grover coin \eqref{2D_grover} and perform the substitution
\begin{align}
\label{substitution}
\theta = \cos{\frac{\Tilde{\theta}}{2}}, \quad \text{i.e.,}\quad
\Tilde{\theta} & = 2 \arccos(\theta),
\end{align}
then the probability distribution $p(x|\theta)$ is the same as in the two cases, and even the wave function is the same. Nevertheless the QFI is extremely different, since in one case there is a divergence and in the other there is not. The origin of such behavior is the Jacobian, $\mathcal{J}$,  of coordinate change

\begin{align}
\label{jacobian}
\frac{\partial}{\partial\theta} = & \mathcal{J} \frac{\partial}{\partial \Tilde{\theta}} = \frac{\partial \Tilde{\theta}}{\partial \theta} \frac{\partial}{\partial \Tilde{\theta}} = \dfrac{-2}{\sqrt{1-\theta^{2}}} \frac{\partial}{\partial \Tilde{\theta}}.
\end{align}
Due to the expression of the QFI for pure states, Eq. \eqref{quantum_fisher_information}, the Jacobian appears as a square modulus as
\begin{equation}
\label{qfi_jacobian}
H_{G}^{(D)}(\theta,t) = \vert \mathcal{J} \vert^2 H_{G}^{(D)}(\Tilde{\theta},t) = \dfrac{4 H_{G}^{(D)}(\Tilde{\theta},t)}{1-\theta^2}.
\end{equation}

The classical FI for pure states shows the same characteristic. The Jacobian appears as a square modulus in Eq. \eqref{fisher_information} and so
\begin{equation}
\label{fi_jacobian}
F_{X; G}^{(D)}(\theta,t) = \vert \mathcal{J} \vert^2 F_{X; G}^{(D)}(\Tilde{\theta},t) = \dfrac{4 F_{X; G}^{(D)}(\Tilde{\theta},t)}{1-\theta^2}.
\end{equation}
while the ratio between FI and QFI is not affected by the change of coordinates and is constant
\begin{equation}
    \label{const_ratio}
    \frac{F_{X; G}^{(D)}(\theta,t)}{H_{G}^{(D)}(\theta,t)}= \textcolor{black}{R(\theta,t)} =\frac{F_{X; G}^{(D)}(\Tilde{\theta},t)}{H_{G}^{(D)}(\Tilde{\theta},t)}\textcolor{black}{=R(\Tilde{\theta},t).}
\end{equation}
We stress that the latter result, Eq. \eqref{const_ratio}, is general and it holds true for any coordinate change, for any dimensionality, and for any coin. Moreover all the considerations made for the rotations, about the FI still hold true for this coin. When we have a family of states that maximize $H_{G}^{(D)}(\theta)$ the optimal state for $F_{X}$ depends both on $\theta$ and $t$.

\section{Summary and conclusions}
\label{conclusion}
We have addressed metrological problems where the parameter of interest, $\theta$, is encoded in the internal degree of freedom of a discrete-time quantum walker, initially localized in position space, and analyzed the precision achievable by different encodings of such parameter. We have shown that coin dimensionality is a potential resource to enhance precision.

When the parameter is encoded in a coin rotation $R_z$, the exact expression of the quantum Fisher information (QFI) has been analytically obtained at any time step $t$. We have determined the initial preparation of the coin state which maximizes the QFI. This optimal state turns out to be independent of the value of the unknown parameter $\theta$. Moreover, the maximum value achievable by the QFI increases with the square of the coin's dimension, according to Eq. \eqref{max_QFI_Rz}. This precision may be achieved by a joint measurement on the two degrees of freedom of the system (walker's position and coin), since in this case the FI associated to a position measurement vanishes.

We have then studied the case where the encoding of the parameter happens through 
$x$- and $y$-rotations, finding the family of initial states that maximize the QFI (also in this they are independent of the value of $\theta$) and showing that the maximal achievable QFI increases with the dimension of the coin. For $D=2$ we have proved the existence of states that are jointly optimal for the two rotation encodings, $R_{x/y}$, and that it is also possible to find states that maximize the QFI for any pair of two-dimensional rotations. For $D=3$, there is an optimal initial state for the two encodings $R_{x}$ and $R_{y}$, which is different 
from that obtained for $R_{z}$. At variance with the $D=2$ case, it is not possible to jointly maximize the QFI of $R_{z}$ and one of the $R_{x/y}$ rotations.

Finally, we have addressed encoding via \textit{generalized Grover coin}, and have found that the optimal states are the same as those optimizing the estimation for $y$-rotation encodings, at least for the dimensions we analyzed ($D=2,3$).

Overall, we have provided evidence that coin dimensionality is a resource to enhance precision in metrological problems involving DTQWs. Our results provide solid tools to address optimization of probe states in several situations of interest, i.e., sensing in magnetic systems, where the coin's degree of freedom is the spin of the particle, or waveguides, where the coin's degree of freedom is the polarization of the photon. 

\begin{acknowledgments}
This work has been done under the auspices of GNFM-INdAM and has been partially 
supported by MUR and EU through the project PRIN22-PNRR-P202222WBL-QWEST.
L.R. acknowledges financial support by INFN through the project ‘QUANTUM’. The main idea was conceived by PB and MGAP. The investigation was carried out by SC and GR and the theoretical framework is due to SC and LR. SC and GR wrote the original draft. Writing review and editing has been performed by LR, SC, PB, and MGAP. 
\end{acknowledgments}

\appendix

\section{Spin Rotation Generators}
\label{app:spin_rotation_gen}
The generators $\mathcal{T}_{n}^{(D)}$ of the rotations about the axes $n = x,y,z$ in dimension $D$ are the representation of the spin operators $S_{n}^{(D)}$ \cite{curtright2014compact}. Here below we report their explicit matrix form.

\subsection{Dimension $D=2$}
In dimension $D=2$ the generators of the rotations are proportional to the Pauli matrices $\{\sigma_n\}_{n=x,y,z}$---generators of $SU(2)$---and read
\begin{align}
&\mathcal{T}_{x}^{(2)}=\frac{\sigma_x}{2}=\frac{1}{2}\begin{pmatrix}
0 & 1 \\
1 & 0
\end{pmatrix} ,\label{tx_2D}\\
&\mathcal{T}_{y}^{(2)}=\frac{\sigma_y}{2}=\frac{1}{2}\begin{pmatrix}
0 & -i \\
i & 0
 \end{pmatrix} , \label{ty_2D}\\
&\mathcal{T}_{z}^{(2)}=\frac{\sigma_z}{2}=\frac{1}{2}\begin{pmatrix}
1 & 0 \\
0 & -1
 \end{pmatrix} .\label{tz_2D}
\end{align}

\subsection{Dimension $D=3$}
In dimension $D=3$ the generators of the rotations are
\begin{equation}
\label{tx_3D}
\mathcal{T}_{x}^{(3)}=\frac{1}{\sqrt{2}}\begin{pmatrix}
0 & 1 & 0 \\
1 & 0 & 1 \\
0 & 1 & 0
 \end{pmatrix} ,
\end{equation}

\begin{equation}
\label{ty_3D}
\mathcal{T}_{y}^{(3)}=\frac{1}{\sqrt{2}}\begin{pmatrix}
0 & -i & 0 \\
i & 0 & -i \\
0 & i & 0
 \end{pmatrix} ,
\end{equation}

\begin{equation}
\label{tz_3D}
\mathcal{T}_{z}^{(3)}=\begin{pmatrix}
1 & 0 & 0 \\
0 & 0 & 0 \\
0 & 0 & -1
 \end{pmatrix} .
\end{equation}

\subsection{Dimension $D=4$}
In dimension $D=4$ the generators of the rotations are
\begin{equation}
\label{tx_4D}
\mathcal{T}_{x}^{(4)}=\frac{1}{2}\begin{pmatrix}
0 & \sqrt{3} & 0 & 0 \\
\sqrt{3} & 0 & 2 & 0 \\
0 & 2 & 0 & \sqrt{3} \\
0 & 0 & \sqrt{3} & 0
 \end{pmatrix} ,
\end{equation}

\begin{equation}
\label{ty_4D}
\mathcal{T}_{y}^{(4)}=\frac{1}{2}\begin{pmatrix}
0 & -i\sqrt{3} & 0 & 0 \\
i\sqrt{3} & 0 & -2i & 0 \\
0 & 2i & 0 & -i\sqrt{3} \\
0 & 0 & i\sqrt{3} & 0
 \end{pmatrix} ,
\end{equation}

\begin{equation}
\label{tz_4D}
\mathcal{T}_{z}^{(4)}=\frac{1}{2}\begin{pmatrix}
3 & 0 & 0 & 0 \\
0 & 1 & 0 & 0 \\
0 & 0 & -1 & 0 \\
0 & 0 & 0 & -3
 \end{pmatrix} .
\end{equation}

\section{Details of some results mentioned in the text}
\label{app:Analytical Calculations}

\subsection{Derivative of the evolution operator}
\label{app:time_evolution}
To assess the QFI as function of time, we need the wave function and its derivative at each time step, i.e., we need their time evolution. While computing $\vert \psi_\theta(t)\rangle = \mathcal{U}^t \vert \psi (0) \rangle$ is straightforward in principle, this might not be the case for $\vert \partial_{\theta} \psi_{\theta} (t) \rangle  \equiv \partial_{\theta} \vert \psi_{\theta} (t) \rangle= (\partial_{\theta} \mathcal{U}^t) \vert \psi (0) \rangle $. To compute the latter we have to compute $\partial_{\theta}\mathcal{U}^t$, as the unknown parameter is encoded in $\mathcal{U}$, not in the initial state, and this operator is the sum of $t$ terms
\begin{equation}
\label{eq:dUt}
\partial_{\theta}\mathcal{U}^t = \sum_{m=0}^{t-1} \mathcal{U}^{m}\left( \partial_{\theta}\mathcal{U} \right) \mathcal{U}^{t-m-1}.
\end{equation}
On the other hand, this expression can be manipulated to iteratively compute such operator as the sum of only two terms at each time step,
\begin{align}
\label{iterative_matrix_derivative}
\partial_{\theta}\mathcal{U}^{t} = \left( \partial_{\theta}\mathcal{U}^{t-1}\right) \mathcal{U} + \mathcal{U}^{t-1} \partial_{\theta} \mathcal{U} \; \; \forall  t > 0.
\end{align}

\subsection{Orthogonality between a state and its derivative}
\label{app:psi_dpsi_orthogonality}
The inner product $\langle  \psi_{\theta} \vert \partial_{\theta} \psi_{\theta} \rangle= i \Im{\langle  \psi_{\theta} \vert \partial_{\theta} \psi_{\theta} \rangle} $ is pure imaginary, as
$\partial_{\theta} \langle \psi_\theta \vert \psi_\theta  \rangle  = 0 =
  2 \Re{\langle \psi_{\theta} \vert \partial_{\theta}\psi_{\theta} \rangle}$
from the normalization condition, $\langle \psi_\theta \vert \psi_\theta  \rangle  = 1$. Then, if the inner product is real, then it is necessarily null, and so the state $\vert \psi_\theta \rangle$ is orthogonal to its derivative $\vert \partial_\theta \psi_\theta\rangle$. This is condition is naturally verified if both the state and its derivative are real.

We want to apply the above result to our estimation problem based on DTQWs. The reason is that the QFI \textit{may} be maximized by making $-\vert \langle  \psi_{\theta} \vert \partial_{\theta} \psi_{\theta} \rangle\vert^2 \leq 0$ in Eq. \eqref{quantum_fisher_information} null. We stress that this argument does not necessarily lead to the true maximum QFI. Nevertheless, it can still be of help in determining the optimal probe. The condition of orthogonality between the wave function $ \vert \psi_{\theta}(t) \rangle  $ and its derivative $\vert \partial_{\theta} \psi_{\theta}(t) \rangle $ at time $t$ is
\begin{align}
\langle  \psi_{\theta}(t) \vert & \partial_{\theta} \psi_{\theta}(t) \rangle  = \langle  \psi(0) \vert (\mathcal{U}^{t})^{\dagger}  (\partial_{\theta}\mathcal{U}^{t}) \vert \psi (0) \rangle \nonumber\\
&= \sum^{t-1}_{m=0} \langle  \psi_{\theta}(m) \vert \mathcal{U}^\dagger ( \partial_{\theta} \mathcal{U}) \vert \psi_{\theta}(m) \rangle \nonumber\\
&= \sum^{t-1}_{m=0} \langle  \psi_{\theta}(m) \vert (\mathbb{1}\otimes \mathcal{C}^\dagger \partial_{\theta} \mathcal{C}) \vert \psi_{\theta}(m) \rangle = 0, \label{eq:orth_cond}
\end{align}
where the second line follows from Eq. \eqref{eq:dUt} and the third line from $\mathcal{S}^\dagger \mathcal{S}=\mathbb{1}$ (unitary). We want this condition to hold true at any time $t>0$, so it must be satisfied also at $t=1$. Given the initial state \eqref{eq:probe_state}, the latter condition reads
\begin{align}
\langle  \psi_{\theta}(1) \vert  \partial_{\theta} \psi_{\theta}(1) \rangle  &= 
 \langle  \psi(0) \vert (\mathbb{1}\otimes \mathcal{C}^\dagger \partial_{\theta} \mathcal{C}) \vert \psi(0) \rangle \nonumber\\
 &= {}_c\langle  \phi^{(D)} \vert \mathcal{C}^\dagger \partial_{\theta} \mathcal{C} \vert \phi^{(D)} \rangle_c = 0 \label{eq:orth_expaval_forall}.
\end{align}
The coin is a unitary operator, thus $\partial_\theta (\mathcal{C}^\dagger \mathcal{C}) = \partial _\theta \mathbb{1} = 0$ which implies
\begin{equation}
(\mathcal{C}^{\dagger} \partial_{\theta} \mathcal{C})^\dagger = - \mathcal{C}^{\dagger} \partial_{\theta} \mathcal{C},
\label{eq:op_antihermitian}
\end{equation}
i.e., the operator $\mathcal{C}^{\dagger} \partial_{\theta} \mathcal{C}$ is anti-Hermitian, so its diagonal elements are pure imaginary. If the condition in Eq. \eqref{eq:orth_expaval_forall} holds true for all $\vert \phi^{(D)} \rangle_c$, then the operator $\mathcal{C}^{\dagger} \partial_{\theta} \mathcal{C}$ is null, which implies that both $\mathcal{C}^{\dagger}$ and $\partial_{\theta} \mathcal{C}$ are singular, unless one of them is null. This cannot be the case as in contradiction with our assumptions: $\mathcal{C}$ does depend on the parameter to be estimated and must be unitary (thus, also non-singular) for $\mathcal{U}$ to be unitary, as required by the DTQW. Therefore, we cannot have a condition holding true for any probe, but in principle we can determine some conditions (sufficient, but not necessary) under which Eq. \eqref{eq:orth_expaval_forall} is satisfied.

The following argument relies on two points: (i) the operator $\mathbb{1} \otimes \mathcal{C}^\dagger \partial_\theta \mathcal{C}$ is anti-Hermitian, because $\mathcal{C}^\dagger \partial_\theta \mathcal{C}$ is, and (ii) the expectation value of an anti-symmetric operator on a real state is null. If the coin operator $\mathcal{C}$ is real, then $\mathcal{C}^\dagger \partial_\theta \mathcal{C}$ and $\mathbb{1} \otimes \mathcal{C}^\dagger \partial_\theta \mathcal{C}$ are real, thus anti-symmetric. As a result, for any real state $\vert \phi^{(D)} \rangle_c$ we have ${}_c\langle  \phi^{(D)} \vert \mathcal{C}^\dagger \partial_{\theta} \mathcal{C} \vert \phi^{(D)} \rangle_c = 0$. The unitary operator $\mathcal{U}$ in Eq. \eqref{time_evolution_operator} is real, being $\mathcal{C}$ real by assumption and $\mathcal{S}$ real by definition, Eq. \eqref{eq:conditional_shift_operator}. If $\vert \phi^{(D)}\rangle_c$ is real, then the initial state \eqref{eq:probe_state} is real and accordingly $\vert \psi_\theta (t) \rangle = \mathcal{U}^t \vert \psi(0)\rangle$ is real for any $t \geq 0$. In conclusion, a sufficient condition to have $\langle  \psi_{\theta}(t) \vert  \partial_{\theta} \psi_{\theta}(t) \rangle = 0$ for all $t>0$ is that the coin operator and the initial coin state are real.

As an example, we discuss the rotations in $D=2$. First, we observe that, for a given the rotation $R_n = \exp(-i \theta \mathcal{T}_n)$---we omit the argument $\theta$ and the superscript $(D)$ for shortness---around the axis $n=x,y,z$, its derivative is $\partial_\theta R_n = -i \mathcal{T}_n  R_n $. Clearly $[R_n,\mathcal{T}_n]=0$ and $R_n^\dagger  R_n = \mathbb{1}$, thus for the rotations we simply have $\mathcal{C}^\dagger \partial_\theta \mathcal{C} = -i \mathcal{T}_n$. From the latter result with the generators in Eqs. \eqref{tx_2D}--\eqref{tz_2D} and by direct inspection of $R_n^{(2)}(\theta)$, we see that only the $y$-rotation it real. The orthogonality condition \eqref{eq:orth_expaval_forall} on the generic initial state \eqref{Bloch_Sphere} for $R_y^{(2)}$ leads to $-\frac{i}{2}\sin\alpha_1 \sin\gamma_1=0$, which is satisfied for $\gamma_1 = 0, \pi$ or $\alpha_1 = 0,\pi$. We numerically verified that the 
resulting probes, i.e
\begin{equation}
\vert \Phi^{(2)}_{y} \rangle_c = \cos{\left( \frac{\alpha_{1}}{2} \right)} \vert -1 \rangle_c \pm \sin{\left( \frac{\alpha_{1}}{2} \right)} \vert +1 \rangle_c,
\end{equation}
are indeed optimal, as they maximize the QFI. On the other hand, the above argument does not apply to $R_{x}^{(2)}(\theta)$ and $R_{z}^{(2)}(\theta)$ due to the presence of 
complex matrices.

\section{Alternative proof of the optimal QFI for $z$-rotations in arbitrary dimension}
\label{app:alt_proof_QFI_max_RzD}
When there is only one parameter $\theta$ to be estimated and the state is pure, the QFI can be alternatively expressed as
\begin{equation}
H(\theta) =\lim_{\delta \theta \to 0}\frac{8 \left( 1-\vert \langle \psi_\theta \vert \psi_{\theta + \delta\theta} \rangle \vert \right )}{\delta\theta^2}.
\label{eq:QFI_pure_1par}
\end{equation}
In our case, the states $\vert \psi_\theta \rangle$ to be considered are reported in Eq. \eqref{eq:psit_RzD} and we can easily prove that
\begin{equation}
\vert \langle \psi_\theta(t) \vert \psi_{\theta+\delta\theta} (t)\rangle \vert = \vert \langle \phi_c  \vert e^{i \delta \theta \mathcal{T}_z^{(D)} t} \vert\phi_c\rangle \vert,
\label{eq:scalar_prod_2Bmin}
\end{equation}
where $\mathcal{T}_z^{(D)}$ is the generator of $z$-rotations in dimension $D$ and $\vert \phi_c\rangle$ is the initial coin state. Maximizing the QFI \eqref{eq:QFI_pure_1par} is equivalent to minimizing Eq. \eqref{eq:scalar_prod_2Bmin}. We define the unitary operator $W\equiv \exp\{i t \mathcal{T}_z^{(D)} \delta \theta\}= \sum_{j=1}^D e^{i\lambda_j}P_j$ whose distinct eigenvalues $\{ e^{i\lambda_j}\}$, with $\lambda_j =  m_j t \delta \theta$, are associated to the corresponding eigenprojectors $\{ P_j\}$. A Lemma by K. R. Parthasarathy \cite{parthasarathy2001consistency} states that, if $\vert \langle \phi_c \vert W \vert \psi_c \rangle \vert >0$ for every normalized state $\vert \phi_c \rangle$, then
\begin{align}
\min_{\norm{\phi_c}=1} \vert \langle \phi_c  \vert W \vert\phi_c\rangle \vert^2 &= \min_{j\neq k}\cos^2\left ( \frac{\lambda_j-\lambda_k}{2}\right )\nonumber\\
&= \cos^2\left ( \frac{\lambda_{j^\ast}-\lambda_{k^\ast}}{2}\right )\nonumber\\
&= \vert \langle \phi_{c}^\ast \vert W \vert \phi_c^\ast \rangle \vert^2,
\label{eq:lemma}
\end{align}
where the eigenvalues $\lambda_{j^\ast}$ and $\lambda_{k^\ast}$ are those minimizing the right-hand side of the first line, and
\begin{align}
\vert \phi_c^\ast \rangle = \frac{1}{\sqrt{2}}(\vert \lambda_{j^\ast} \rangle + \vert \lambda_{k^\ast} \rangle ),
\label{eq:lemma_optim_state}
\end{align}
where $\vert \lambda_{j^\ast} \rangle$ and $\vert \lambda_{k^\ast} \rangle$ are arbitrary normalized states in the range of $P_{j^\ast}$ and $P_{k^\ast}$, respectively.

According to this lemma, we can prove the maximum QFI \eqref{max_QFI_Rz} and the optimal initial coin state \eqref{initial_max_QFI_Rz}. Since the QFI is defined in the limit for $\delta \theta \to 0$, we can compute the square root of Eq. \eqref{eq:lemma} as
\begin{gather}
    \sqrt{ \min_{j \neq k}\cos^2\left ( \frac{\lambda_{j}-\lambda_{k }}{2}\right )} \approx \sqrt{1-\frac{t^2 \delta \theta^2}{4}\max_{j \neq k}(m_j - m_k)^2}\nonumber\\
    =\sqrt{1- t^2 s^2 \delta \theta^2}
    \approx 1- \frac{t^2 s^2}{2} \delta \theta^2,
\end{gather}
because $-s \leq m_j \leq s$. This, together with Eq. \eqref{eq:QFI_pure_1par}, leads to Eq. \eqref{max_QFI_Rz}. Now, we focus on the optimal probe state. In our case, all the $D$ eigenvalues of $\mathcal{T}^{(D)}_z$, and so of $W$, are distinct. Hence, an operator $P_{j}$ is a projector onto the one-dimensional space spanned by the $j$-th eigenstate of $\mathcal{T}^{(D)}_z$. As shown above, the two eigenstates we need to define the state in Eq. \eqref{eq:lemma_optim_state} correspond to the lowest and highest eigenvalues of $\mathcal{T}^{(D)}_z$, i.e., we need $\vert \pm  s\rangle$. Each eigenspace is one-dimensional, thus the only arbitrariness left is in the choice of the phase factor. However, when taking the superposition of two states, only the relative phase matters. This, together with Eq. \eqref{eq:lemma_optim_state}, leads to the optimal initial coin state in Eq. \eqref{initial_max_QFI_Rz}.

\bibliography{cdbiblio.bib}

\begin{thebibliography}{62}%
\makeatletter
\providecommand \@ifxundefined [1]{%
 \@ifx{#1\undefined}
}%
\providecommand \@ifnum [1]{%
 \ifnum #1\expandafter \@firstoftwo
 \else \expandafter \@secondoftwo
 \fi
}%
\providecommand \@ifx [1]{%
 \ifx #1\expandafter \@firstoftwo
 \else \expandafter \@secondoftwo
 \fi
}%
\providecommand \natexlab [1]{#1}%
\providecommand \enquote  [1]{``#1''}%
\providecommand \bibnamefont  [1]{#1}%
\providecommand \bibfnamefont [1]{#1}%
\providecommand \citenamefont [1]{#1}%
\providecommand \href@noop [0]{\@secondoftwo}%
\providecommand \href [0]{\begingroup \@sanitize@url \@href}%
\providecommand \@href[1]{\@@startlink{#1}\@@href}%
\providecommand \@@href[1]{\endgroup#1\@@endlink}%
\providecommand \@sanitize@url [0]{\catcode `\\12\catcode `\$12\catcode
  `\&12\catcode `\#12\catcode `\^12\catcode `\_12\catcode `\%12\relax}%
\providecommand \@@startlink[1]{}%
\providecommand \@@endlink[0]{}%
\providecommand \url  [0]{\begingroup\@sanitize@url \@url }%
\providecommand \@url [1]{\endgroup\@href {#1}{\urlprefix }}%
\providecommand \urlprefix  [0]{URL }%
\providecommand \Eprint [0]{\href }%
\providecommand \doibase [0]{https://doi.org/}%
\providecommand \selectlanguage [0]{\@gobble}%
\providecommand \bibinfo  [0]{\@secondoftwo}%
\providecommand \bibfield  [0]{\@secondoftwo}%
\providecommand \translation [1]{[#1]}%
\providecommand \BibitemOpen [0]{}%
\providecommand \bibitemStop [0]{}%
\providecommand \bibitemNoStop [0]{.\EOS\space}%
\providecommand \EOS [0]{\spacefactor3000\relax}%
\providecommand \BibitemShut  [1]{\csname bibitem#1\endcsname}%
\let\auto@bib@innerbib\@empty
\bibitem [{\citenamefont {Wang}\ and\ \citenamefont
  {Manouchehri}(2013)}]{wang2013physical}%
  \BibitemOpen
  \bibfield  {author} {\bibinfo {author} {\bibfnamefont {J.}~\bibnamefont
  {Wang}}\ and\ \bibinfo {author} {\bibfnamefont {K.}~\bibnamefont
  {Manouchehri}},\ }\href@noop {} {\emph {\bibinfo {title} {Physical
  Implementation of Quantum Walks}}}\ (\bibinfo  {publisher} {Springer},\
  \bibinfo {address} {New York},\ \bibinfo {year} {2013})\BibitemShut {NoStop}%
\bibitem [{\citenamefont {Portugal}(2018)}]{portugal2018quantum}%
  \BibitemOpen
  \bibfield  {author} {\bibinfo {author} {\bibfnamefont {R.}~\bibnamefont
  {Portugal}},\ }\href@noop {} {\emph {\bibinfo {title} {Quantum Walks and
  Search Algorithms}}}\ (\bibinfo  {publisher} {Springer},\ \bibinfo {address}
  {New York},\ \bibinfo {year} {2018})\BibitemShut {NoStop}%
\bibitem [{\citenamefont {Kadian}\ \emph {et~al.}(2021)\citenamefont {Kadian},
  \citenamefont {Garhwal},\ and\ \citenamefont {Kumar}}]{KADIAN2021100419}%
  \BibitemOpen
  \bibfield  {author} {\bibinfo {author} {\bibfnamefont {K.}~\bibnamefont
  {Kadian}}, \bibinfo {author} {\bibfnamefont {S.}~\bibnamefont {Garhwal}},\
  and\ \bibinfo {author} {\bibfnamefont {A.}~\bibnamefont {Kumar}},\ }\href
  {https://doi.org/https://doi.org/10.1016/j.cosrev.2021.100419} {\bibfield
  {journal} {\bibinfo  {journal} {Computer Science Review}\ }\textbf {\bibinfo
  {volume} {41}},\ \bibinfo {pages} {100419} (\bibinfo {year}
  {2021})}\BibitemShut {NoStop}%
\bibitem [{\citenamefont {Shenvi}\ \emph {et~al.}(2003)\citenamefont {Shenvi},
  \citenamefont {Kempe},\ and\ \citenamefont {Whaley}}]{PhysRevA.67.052307}%
  \BibitemOpen
  \bibfield  {author} {\bibinfo {author} {\bibfnamefont {N.}~\bibnamefont
  {Shenvi}}, \bibinfo {author} {\bibfnamefont {J.}~\bibnamefont {Kempe}},\ and\
  \bibinfo {author} {\bibfnamefont {K.~B.}\ \bibnamefont {Whaley}},\ }\href
  {https://doi.org/10.1103/PhysRevA.67.052307} {\bibfield  {journal} {\bibinfo
  {journal} {Phys. Rev. A}\ }\textbf {\bibinfo {volume} {67}},\ \bibinfo
  {pages} {052307} (\bibinfo {year} {2003})}\BibitemShut {NoStop}%
\bibitem [{\citenamefont {Ambainis}(2003)}]{ambainis2003quantum}%
  \BibitemOpen
  \bibfield  {author} {\bibinfo {author} {\bibfnamefont {A.}~\bibnamefont
  {Ambainis}},\ }\href@noop {} {\bibfield  {journal} {\bibinfo  {journal}
  {International Journal of Quantum Information}\ }\textbf {\bibinfo {volume}
  {1}},\ \bibinfo {pages} {507} (\bibinfo {year} {2003})}\BibitemShut {NoStop}%
\bibitem [{\citenamefont {Lovett}\ \emph {et~al.}(2010)\citenamefont {Lovett},
  \citenamefont {Cooper}, \citenamefont {Everitt}, \citenamefont {Trevers},\
  and\ \citenamefont {Kendon}}]{lovett2010universal}%
  \BibitemOpen
  \bibfield  {author} {\bibinfo {author} {\bibfnamefont {N.~B.}\ \bibnamefont
  {Lovett}}, \bibinfo {author} {\bibfnamefont {S.}~\bibnamefont {Cooper}},
  \bibinfo {author} {\bibfnamefont {M.}~\bibnamefont {Everitt}}, \bibinfo
  {author} {\bibfnamefont {M.}~\bibnamefont {Trevers}},\ and\ \bibinfo {author}
  {\bibfnamefont {V.}~\bibnamefont {Kendon}},\ }\href
  {https://doi.org/10.1103/PhysRevA.81.042330} {\bibfield  {journal} {\bibinfo
  {journal} {Phys. Rev. A}\ }\textbf {\bibinfo {volume} {81}},\ \bibinfo
  {pages} {042330} (\bibinfo {year} {2010})}\BibitemShut {NoStop}%
\bibitem [{\citenamefont {Singh}\ \emph {et~al.}(2021)\citenamefont {Singh},
  \citenamefont {Chawla}, \citenamefont {Sarkar},\ and\ \citenamefont
  {Chandrashekar}}]{singh2021universal}%
  \BibitemOpen
  \bibfield  {author} {\bibinfo {author} {\bibfnamefont {S.}~\bibnamefont
  {Singh}}, \bibinfo {author} {\bibfnamefont {P.}~\bibnamefont {Chawla}},
  \bibinfo {author} {\bibfnamefont {A.}~\bibnamefont {Sarkar}},\ and\ \bibinfo
  {author} {\bibfnamefont {C.}~\bibnamefont {Chandrashekar}},\ }\href
  {https://doi.org/10.1038/s41598-021-91033-5} {\bibfield  {journal} {\bibinfo
  {journal} {Scientific Reports}\ }\textbf {\bibinfo {volume} {11}},\ \bibinfo
  {pages} {11551} (\bibinfo {year} {2021})}\BibitemShut {NoStop}%
\bibitem [{\citenamefont {Chatterjee}\ \emph {et~al.}(2020)\citenamefont
  {Chatterjee}, \citenamefont {Devrari}, \citenamefont {Behera},\ and\
  \citenamefont {Panigrahi}}]{chatterjee2020experimental}%
  \BibitemOpen
  \bibfield  {author} {\bibinfo {author} {\bibfnamefont {Y.}~\bibnamefont
  {Chatterjee}}, \bibinfo {author} {\bibfnamefont {V.}~\bibnamefont {Devrari}},
  \bibinfo {author} {\bibfnamefont {B.~K.}\ \bibnamefont {Behera}},\ and\
  \bibinfo {author} {\bibfnamefont {P.~K.}\ \bibnamefont {Panigrahi}},\
  }\href@noop {} {\bibfield  {journal} {\bibinfo  {journal} {Quantum
  Information Processing}\ }\textbf {\bibinfo {volume} {19}},\ \bibinfo {pages}
  {1} (\bibinfo {year} {2020})}\BibitemShut {NoStop}%
\bibitem [{\citenamefont {Abd-El-Atty}\ \emph {et~al.}(2021)\citenamefont
  {Abd-El-Atty}, \citenamefont {Iliyasu}, \citenamefont {Alanezi},\ and\
  \citenamefont {Abd El-latif}}]{abd2021optical}%
  \BibitemOpen
  \bibfield  {author} {\bibinfo {author} {\bibfnamefont {B.}~\bibnamefont
  {Abd-El-Atty}}, \bibinfo {author} {\bibfnamefont {A.~M.}\ \bibnamefont
  {Iliyasu}}, \bibinfo {author} {\bibfnamefont {A.}~\bibnamefont {Alanezi}},\
  and\ \bibinfo {author} {\bibfnamefont {A.~A.}\ \bibnamefont {Abd El-latif}},\
  }\href@noop {} {\bibfield  {journal} {\bibinfo  {journal} {Optics and Lasers
  in Engineering}\ }\textbf {\bibinfo {volume} {138}},\ \bibinfo {pages}
  {106403} (\bibinfo {year} {2021})}\BibitemShut {NoStop}%
\bibitem [{\citenamefont {Abd El-Latif}\ \emph {et~al.}(2020)\citenamefont {Abd
  El-Latif}, \citenamefont {Abd-El-Atty}, \citenamefont {Mazurczyk},
  \citenamefont {Fung},\ and\ \citenamefont {Venegas-Andraca}}]{abd2020secure}%
  \BibitemOpen
  \bibfield  {author} {\bibinfo {author} {\bibfnamefont {A.~A.}\ \bibnamefont
  {Abd El-Latif}}, \bibinfo {author} {\bibfnamefont {B.}~\bibnamefont
  {Abd-El-Atty}}, \bibinfo {author} {\bibfnamefont {W.}~\bibnamefont
  {Mazurczyk}}, \bibinfo {author} {\bibfnamefont {C.}~\bibnamefont {Fung}},\
  and\ \bibinfo {author} {\bibfnamefont {S.~E.}\ \bibnamefont
  {Venegas-Andraca}},\ }\href@noop {} {\bibfield  {journal} {\bibinfo
  {journal} {IEEE Transactions on Network and Service Management}\ }\textbf
  {\bibinfo {volume} {17}},\ \bibinfo {pages} {118} (\bibinfo {year}
  {2020})}\BibitemShut {NoStop}%
\bibitem [{\citenamefont {Liang}\ \emph {et~al.}(2022)\citenamefont {Liang},
  \citenamefont {Yan}, \citenamefont {Iliyasu}, \citenamefont {Salama},\ and\
  \citenamefont {Hirota}}]{liang2022hadamard}%
  \BibitemOpen
  \bibfield  {author} {\bibinfo {author} {\bibfnamefont {W.}~\bibnamefont
  {Liang}}, \bibinfo {author} {\bibfnamefont {F.}~\bibnamefont {Yan}}, \bibinfo
  {author} {\bibfnamefont {A.~M.}\ \bibnamefont {Iliyasu}}, \bibinfo {author}
  {\bibfnamefont {A.~S.}\ \bibnamefont {Salama}},\ and\ \bibinfo {author}
  {\bibfnamefont {K.}~\bibnamefont {Hirota}},\ }\href@noop {} {\bibfield
  {journal} {\bibinfo  {journal} {Computer Communications}\ }\textbf {\bibinfo
  {volume} {193}},\ \bibinfo {pages} {378} (\bibinfo {year}
  {2022})}\BibitemShut {NoStop}%
\bibitem [{\citenamefont {Travaglione}\ and\ \citenamefont
  {Milburn}(2002)}]{travaglione2002implementing}%
  \BibitemOpen
  \bibfield  {author} {\bibinfo {author} {\bibfnamefont {B.~C.}\ \bibnamefont
  {Travaglione}}\ and\ \bibinfo {author} {\bibfnamefont {G.~J.}\ \bibnamefont
  {Milburn}},\ }\href@noop {} {\bibfield  {journal} {\bibinfo  {journal}
  {Physical Review A}\ }\textbf {\bibinfo {volume} {65}},\ \bibinfo {pages}
  {032310} (\bibinfo {year} {2002})}\BibitemShut {NoStop}%
\bibitem [{\citenamefont {D{\"u}r}\ \emph {et~al.}(2002)\citenamefont
  {D{\"u}r}, \citenamefont {Raussendorf}, \citenamefont {Kendon},\ and\
  \citenamefont {Briegel}}]{dur2002quantum}%
  \BibitemOpen
  \bibfield  {author} {\bibinfo {author} {\bibfnamefont {W.}~\bibnamefont
  {D{\"u}r}}, \bibinfo {author} {\bibfnamefont {R.}~\bibnamefont
  {Raussendorf}}, \bibinfo {author} {\bibfnamefont {V.~M.}\ \bibnamefont
  {Kendon}},\ and\ \bibinfo {author} {\bibfnamefont {H.-J.}\ \bibnamefont
  {Briegel}},\ }\href@noop {} {\bibfield  {journal} {\bibinfo  {journal}
  {Physical Review A}\ }\textbf {\bibinfo {volume} {66}},\ \bibinfo {pages}
  {052319} (\bibinfo {year} {2002})}\BibitemShut {NoStop}%
\bibitem [{\citenamefont {Karski}\ \emph {et~al.}(2009)\citenamefont {Karski},
  \citenamefont {F{\"o}rster}, \citenamefont {Choi}, \citenamefont {Steffen},
  \citenamefont {Alt}, \citenamefont {Meschede},\ and\ \citenamefont
  {Widera}}]{karski2009quantum}%
  \BibitemOpen
  \bibfield  {author} {\bibinfo {author} {\bibfnamefont {M.}~\bibnamefont
  {Karski}}, \bibinfo {author} {\bibfnamefont {L.}~\bibnamefont {F{\"o}rster}},
  \bibinfo {author} {\bibfnamefont {J.-M.}\ \bibnamefont {Choi}}, \bibinfo
  {author} {\bibfnamefont {A.}~\bibnamefont {Steffen}}, \bibinfo {author}
  {\bibfnamefont {W.}~\bibnamefont {Alt}}, \bibinfo {author} {\bibfnamefont
  {D.}~\bibnamefont {Meschede}},\ and\ \bibinfo {author} {\bibfnamefont
  {A.}~\bibnamefont {Widera}},\ }\href@noop {} {\bibfield  {journal} {\bibinfo
  {journal} {Science}\ }\textbf {\bibinfo {volume} {325}},\ \bibinfo {pages}
  {174} (\bibinfo {year} {2009})}\BibitemShut {NoStop}%
\bibitem [{\citenamefont {Ehrhardt}\ \emph {et~al.}(2021)\citenamefont
  {Ehrhardt}, \citenamefont {Keil}, \citenamefont {Maczewsky}, \citenamefont
  {Dittel}, \citenamefont {Heinrich},\ and\ \citenamefont
  {Szameit}}]{ehrhardt2021exploring}%
  \BibitemOpen
  \bibfield  {author} {\bibinfo {author} {\bibfnamefont {M.}~\bibnamefont
  {Ehrhardt}}, \bibinfo {author} {\bibfnamefont {R.}~\bibnamefont {Keil}},
  \bibinfo {author} {\bibfnamefont {L.~J.}\ \bibnamefont {Maczewsky}}, \bibinfo
  {author} {\bibfnamefont {C.}~\bibnamefont {Dittel}}, \bibinfo {author}
  {\bibfnamefont {M.}~\bibnamefont {Heinrich}},\ and\ \bibinfo {author}
  {\bibfnamefont {A.}~\bibnamefont {Szameit}},\ }\href@noop {} {\bibfield
  {journal} {\bibinfo  {journal} {Science Advances}\ }\textbf {\bibinfo
  {volume} {7}},\ \bibinfo {pages} {eabc5266} (\bibinfo {year}
  {2021})}\BibitemShut {NoStop}%
\bibitem [{\citenamefont {Perets}\ \emph {et~al.}(2008)\citenamefont {Perets},
  \citenamefont {Lahini}, \citenamefont {Pozzi}, \citenamefont {Sorel},
  \citenamefont {Morandotti},\ and\ \citenamefont
  {Silberberg}}]{perets2008realization}%
  \BibitemOpen
  \bibfield  {author} {\bibinfo {author} {\bibfnamefont {H.~B.}\ \bibnamefont
  {Perets}}, \bibinfo {author} {\bibfnamefont {Y.}~\bibnamefont {Lahini}},
  \bibinfo {author} {\bibfnamefont {F.}~\bibnamefont {Pozzi}}, \bibinfo
  {author} {\bibfnamefont {M.}~\bibnamefont {Sorel}}, \bibinfo {author}
  {\bibfnamefont {R.}~\bibnamefont {Morandotti}},\ and\ \bibinfo {author}
  {\bibfnamefont {Y.}~\bibnamefont {Silberberg}},\ }\href@noop {} {\bibfield
  {journal} {\bibinfo  {journal} {Physical review letters}\ }\textbf {\bibinfo
  {volume} {100}},\ \bibinfo {pages} {170506} (\bibinfo {year}
  {2008})}\BibitemShut {NoStop}%
\bibitem [{\citenamefont {Broome}\ \emph {et~al.}(2010)\citenamefont {Broome},
  \citenamefont {Fedrizzi}, \citenamefont {Lanyon}, \citenamefont {Kassal},
  \citenamefont {Aspuru-Guzik},\ and\ \citenamefont
  {White}}]{broome2010discrete}%
  \BibitemOpen
  \bibfield  {author} {\bibinfo {author} {\bibfnamefont {M.~A.}\ \bibnamefont
  {Broome}}, \bibinfo {author} {\bibfnamefont {A.}~\bibnamefont {Fedrizzi}},
  \bibinfo {author} {\bibfnamefont {B.~P.}\ \bibnamefont {Lanyon}}, \bibinfo
  {author} {\bibfnamefont {I.}~\bibnamefont {Kassal}}, \bibinfo {author}
  {\bibfnamefont {A.}~\bibnamefont {Aspuru-Guzik}},\ and\ \bibinfo {author}
  {\bibfnamefont {A.~G.}\ \bibnamefont {White}},\ }\href@noop {} {\bibfield
  {journal} {\bibinfo  {journal} {Physical review letters}\ }\textbf {\bibinfo
  {volume} {104}},\ \bibinfo {pages} {153602} (\bibinfo {year}
  {2010})}\BibitemShut {NoStop}%
\bibitem [{\citenamefont {Qiang}\ \emph {et~al.}(2021)\citenamefont {Qiang},
  \citenamefont {Wang}, \citenamefont {Xue}, \citenamefont {Ge}, \citenamefont
  {Chen}, \citenamefont {Liu}, \citenamefont {Huang}, \citenamefont {Fu},
  \citenamefont {Xu}, \citenamefont {Yi} \emph
  {et~al.}}]{qiang2021implementing}%
  \BibitemOpen
  \bibfield  {author} {\bibinfo {author} {\bibfnamefont {X.}~\bibnamefont
  {Qiang}}, \bibinfo {author} {\bibfnamefont {Y.}~\bibnamefont {Wang}},
  \bibinfo {author} {\bibfnamefont {S.}~\bibnamefont {Xue}}, \bibinfo {author}
  {\bibfnamefont {R.}~\bibnamefont {Ge}}, \bibinfo {author} {\bibfnamefont
  {L.}~\bibnamefont {Chen}}, \bibinfo {author} {\bibfnamefont {Y.}~\bibnamefont
  {Liu}}, \bibinfo {author} {\bibfnamefont {A.}~\bibnamefont {Huang}}, \bibinfo
  {author} {\bibfnamefont {X.}~\bibnamefont {Fu}}, \bibinfo {author}
  {\bibfnamefont {P.}~\bibnamefont {Xu}}, \bibinfo {author} {\bibfnamefont
  {T.}~\bibnamefont {Yi}}, \emph {et~al.},\ }\href@noop {} {\bibfield
  {journal} {\bibinfo  {journal} {Science Advances}\ }\textbf {\bibinfo
  {volume} {7}},\ \bibinfo {pages} {eabb8375} (\bibinfo {year}
  {2021})}\BibitemShut {NoStop}%
\bibitem [{\citenamefont {Tang}\ \emph {et~al.}(2020)\citenamefont {Tang},
  \citenamefont {Hou}, \citenamefont {Shang}, \citenamefont {Zhu},
  \citenamefont {Xiang}, \citenamefont {Li},\ and\ \citenamefont
  {Guo}}]{tang2020experimental}%
  \BibitemOpen
  \bibfield  {author} {\bibinfo {author} {\bibfnamefont {J.-F.}\ \bibnamefont
  {Tang}}, \bibinfo {author} {\bibfnamefont {Z.}~\bibnamefont {Hou}}, \bibinfo
  {author} {\bibfnamefont {J.}~\bibnamefont {Shang}}, \bibinfo {author}
  {\bibfnamefont {H.}~\bibnamefont {Zhu}}, \bibinfo {author} {\bibfnamefont
  {G.-Y.}\ \bibnamefont {Xiang}}, \bibinfo {author} {\bibfnamefont {C.-F.}\
  \bibnamefont {Li}},\ and\ \bibinfo {author} {\bibfnamefont {G.-C.}\
  \bibnamefont {Guo}},\ }\href@noop {} {\bibfield  {journal} {\bibinfo
  {journal} {Physical Review Letters}\ }\textbf {\bibinfo {volume} {124}},\
  \bibinfo {pages} {060502} (\bibinfo {year} {2020})}\BibitemShut {NoStop}%
\bibitem [{\citenamefont {Di~Colandrea}\ \emph {et~al.}(2023)\citenamefont
  {Di~Colandrea}, \citenamefont {Babazadeh}, \citenamefont {Dauphin},
  \citenamefont {Massignan}, \citenamefont {Marrucci},\ and\ \citenamefont
  {Cardano}}]{di2023ultra}%
  \BibitemOpen
  \bibfield  {author} {\bibinfo {author} {\bibfnamefont {F.}~\bibnamefont
  {Di~Colandrea}}, \bibinfo {author} {\bibfnamefont {A.}~\bibnamefont
  {Babazadeh}}, \bibinfo {author} {\bibfnamefont {A.}~\bibnamefont {Dauphin}},
  \bibinfo {author} {\bibfnamefont {P.}~\bibnamefont {Massignan}}, \bibinfo
  {author} {\bibfnamefont {L.}~\bibnamefont {Marrucci}},\ and\ \bibinfo
  {author} {\bibfnamefont {F.}~\bibnamefont {Cardano}},\ }\href@noop {}
  {\bibfield  {journal} {\bibinfo  {journal} {Optica}\ }\textbf {\bibinfo
  {volume} {10}},\ \bibinfo {pages} {324} (\bibinfo {year} {2023})}\BibitemShut
  {NoStop}%
\bibitem [{\citenamefont {Xie}\ \emph {et~al.}(2020)\citenamefont {Xie},
  \citenamefont {Deng}, \citenamefont {Xiao}, \citenamefont {Gou},
  \citenamefont {Chen}, \citenamefont {Yi},\ and\ \citenamefont
  {Yan}}]{xie2020topological}%
  \BibitemOpen
  \bibfield  {author} {\bibinfo {author} {\bibfnamefont {D.}~\bibnamefont
  {Xie}}, \bibinfo {author} {\bibfnamefont {T.-S.}\ \bibnamefont {Deng}},
  \bibinfo {author} {\bibfnamefont {T.}~\bibnamefont {Xiao}}, \bibinfo {author}
  {\bibfnamefont {W.}~\bibnamefont {Gou}}, \bibinfo {author} {\bibfnamefont
  {T.}~\bibnamefont {Chen}}, \bibinfo {author} {\bibfnamefont {W.}~\bibnamefont
  {Yi}},\ and\ \bibinfo {author} {\bibfnamefont {B.}~\bibnamefont {Yan}},\
  }\href@noop {} {\bibfield  {journal} {\bibinfo  {journal} {Physical Review
  Letters}\ }\textbf {\bibinfo {volume} {124}},\ \bibinfo {pages} {050502}
  (\bibinfo {year} {2020})}\BibitemShut {NoStop}%
\bibitem [{\citenamefont {Esposito}\ \emph {et~al.}(2022)\citenamefont
  {Esposito}, \citenamefont {Barros}, \citenamefont {Dur{\'a}n~Hern{\'a}ndez},
  \citenamefont {Carvacho}, \citenamefont {Di~Colandrea}, \citenamefont
  {Barboza}, \citenamefont {Cardano}, \citenamefont {Spagnolo}, \citenamefont
  {Marrucci},\ and\ \citenamefont {Sciarrino}}]{esposito2022quantum}%
  \BibitemOpen
  \bibfield  {author} {\bibinfo {author} {\bibfnamefont {C.}~\bibnamefont
  {Esposito}}, \bibinfo {author} {\bibfnamefont {M.~R.}\ \bibnamefont
  {Barros}}, \bibinfo {author} {\bibfnamefont {A.}~\bibnamefont
  {Dur{\'a}n~Hern{\'a}ndez}}, \bibinfo {author} {\bibfnamefont
  {G.}~\bibnamefont {Carvacho}}, \bibinfo {author} {\bibfnamefont
  {F.}~\bibnamefont {Di~Colandrea}}, \bibinfo {author} {\bibfnamefont
  {R.}~\bibnamefont {Barboza}}, \bibinfo {author} {\bibfnamefont
  {F.}~\bibnamefont {Cardano}}, \bibinfo {author} {\bibfnamefont
  {N.}~\bibnamefont {Spagnolo}}, \bibinfo {author} {\bibfnamefont
  {L.}~\bibnamefont {Marrucci}},\ and\ \bibinfo {author} {\bibfnamefont
  {F.}~\bibnamefont {Sciarrino}},\ }\href@noop {} {\bibfield  {journal}
  {\bibinfo  {journal} {npj Quantum Information}\ }\textbf {\bibinfo {volume}
  {8}},\ \bibinfo {pages} {34} (\bibinfo {year} {2022})}\BibitemShut {NoStop}%
\bibitem [{\citenamefont {Acasiete}\ \emph {et~al.}(2020)\citenamefont
  {Acasiete}, \citenamefont {Agostini}, \citenamefont {Moqadam},\ and\
  \citenamefont {Portugal}}]{acasiete2020implementation}%
  \BibitemOpen
  \bibfield  {author} {\bibinfo {author} {\bibfnamefont {F.}~\bibnamefont
  {Acasiete}}, \bibinfo {author} {\bibfnamefont {F.~P.}\ \bibnamefont
  {Agostini}}, \bibinfo {author} {\bibfnamefont {J.~K.}\ \bibnamefont
  {Moqadam}},\ and\ \bibinfo {author} {\bibfnamefont {R.}~\bibnamefont
  {Portugal}},\ }\href@noop {} {\bibfield  {journal} {\bibinfo  {journal}
  {Quantum Information Processing}\ }\textbf {\bibinfo {volume} {19}},\
  \bibinfo {pages} {1} (\bibinfo {year} {2020})}\BibitemShut {NoStop}%
\bibitem [{\citenamefont {Romanelli}(2009)}]{PhysRevA.80.042332}%
  \BibitemOpen
  \bibfield  {author} {\bibinfo {author} {\bibfnamefont {A.}~\bibnamefont
  {Romanelli}},\ }\href {https://doi.org/10.1103/PhysRevA.80.042332} {\bibfield
   {journal} {\bibinfo  {journal} {Phys. Rev. A}\ }\textbf {\bibinfo {volume}
  {80}},\ \bibinfo {pages} {042332} (\bibinfo {year} {2009})}\BibitemShut
  {NoStop}%
\bibitem [{\citenamefont {Walczak}\ and\ \citenamefont
  {Bauer}(2021)}]{PhysRevE.104.064209}%
  \BibitemOpen
  \bibfield  {author} {\bibinfo {author} {\bibfnamefont {Z.}~\bibnamefont
  {Walczak}}\ and\ \bibinfo {author} {\bibfnamefont {J.~H.}\ \bibnamefont
  {Bauer}},\ }\href {https://doi.org/10.1103/PhysRevE.104.064209} {\bibfield
  {journal} {\bibinfo  {journal} {Phys. Rev. E}\ }\textbf {\bibinfo {volume}
  {104}},\ \bibinfo {pages} {064209} (\bibinfo {year} {2021})}\BibitemShut
  {NoStop}%
\bibitem [{\citenamefont {Vieira}\ \emph {et~al.}(2013)\citenamefont {Vieira},
  \citenamefont {Amorim},\ and\ \citenamefont
  {Rigolin}}]{PhysRevLett.111.180503}%
  \BibitemOpen
  \bibfield  {author} {\bibinfo {author} {\bibfnamefont {R.}~\bibnamefont
  {Vieira}}, \bibinfo {author} {\bibfnamefont {E.~P.~M.}\ \bibnamefont
  {Amorim}},\ and\ \bibinfo {author} {\bibfnamefont {G.}~\bibnamefont
  {Rigolin}},\ }\href {https://doi.org/10.1103/PhysRevLett.111.180503}
  {\bibfield  {journal} {\bibinfo  {journal} {Phys. Rev. Lett.}\ }\textbf
  {\bibinfo {volume} {111}},\ \bibinfo {pages} {180503} (\bibinfo {year}
  {2013})}\BibitemShut {NoStop}%
\bibitem [{\citenamefont {Kendon}(2006)}]{kendon2006}%
  \BibitemOpen
  \bibfield  {author} {\bibinfo {author} {\bibfnamefont {V.}~\bibnamefont
  {Kendon}},\ }\href {https://doi.org/10.1142/S0219749906002195} {\bibfield
  {journal} {\bibinfo  {journal} {International Journal of Quantum
  Information}\ }\textbf {\bibinfo {volume} {04}},\ \bibinfo {pages} {791}
  (\bibinfo {year} {2006})}\BibitemShut {NoStop}%
\bibitem [{\citenamefont {Goyal}\ \emph {et~al.}(2015)\citenamefont {Goyal},
  \citenamefont {Roux}, \citenamefont {Forbes},\ and\ \citenamefont
  {Konrad}}]{PhysRevA.92.040302}%
  \BibitemOpen
  \bibfield  {author} {\bibinfo {author} {\bibfnamefont {S.~K.}\ \bibnamefont
  {Goyal}}, \bibinfo {author} {\bibfnamefont {F.~S.}\ \bibnamefont {Roux}},
  \bibinfo {author} {\bibfnamefont {A.}~\bibnamefont {Forbes}},\ and\ \bibinfo
  {author} {\bibfnamefont {T.}~\bibnamefont {Konrad}},\ }\href
  {https://doi.org/10.1103/PhysRevA.92.040302} {\bibfield  {journal} {\bibinfo
  {journal} {Phys. Rev. A}\ }\textbf {\bibinfo {volume} {92}},\ \bibinfo
  {pages} {040302} (\bibinfo {year} {2015})}\BibitemShut {NoStop}%
\bibitem [{\citenamefont {Mukai}\ and\ \citenamefont
  {Hatano}(2020)}]{PhysRevResearch.2.023378}%
  \BibitemOpen
  \bibfield  {author} {\bibinfo {author} {\bibfnamefont {K.}~\bibnamefont
  {Mukai}}\ and\ \bibinfo {author} {\bibfnamefont {N.}~\bibnamefont {Hatano}},\
  }\href {https://doi.org/10.1103/PhysRevResearch.2.023378} {\bibfield
  {journal} {\bibinfo  {journal} {Phys. Rev. Res.}\ }\textbf {\bibinfo {volume}
  {2}},\ \bibinfo {pages} {023378} (\bibinfo {year} {2020})}\BibitemShut
  {NoStop}%
\bibitem [{\citenamefont {Brun}\ \emph {et~al.}(2003)\citenamefont {Brun},
  \citenamefont {Carteret},\ and\ \citenamefont
  {Ambainis}}]{PhysRevA.67.052317}%
  \BibitemOpen
  \bibfield  {author} {\bibinfo {author} {\bibfnamefont {T.~A.}\ \bibnamefont
  {Brun}}, \bibinfo {author} {\bibfnamefont {H.~A.}\ \bibnamefont {Carteret}},\
  and\ \bibinfo {author} {\bibfnamefont {A.}~\bibnamefont {Ambainis}},\ }\href
  {https://doi.org/10.1103/PhysRevA.67.052317} {\bibfield  {journal} {\bibinfo
  {journal} {Phys. Rev. A}\ }\textbf {\bibinfo {volume} {67}},\ \bibinfo
  {pages} {052317} (\bibinfo {year} {2003})}\BibitemShut {NoStop}%
\bibitem [{\citenamefont {Segawa}\ and\ \citenamefont
  {Konno}(2008)}]{segawa2008}%
  \BibitemOpen
  \bibfield  {author} {\bibinfo {author} {\bibfnamefont {E.}~\bibnamefont
  {Segawa}}\ and\ \bibinfo {author} {\bibfnamefont {N.}~\bibnamefont {Konno}},\
  }\href {https://doi.org/10.1142/S0219749908004456} {\bibfield  {journal}
  {\bibinfo  {journal} {International Journal of Quantum Information}\ }\textbf
  {\bibinfo {volume} {06}},\ \bibinfo {pages} {1231} (\bibinfo {year}
  {2008})}\BibitemShut {NoStop}%
\bibitem [{\citenamefont {Shang}\ \emph {et~al.}(2019)\citenamefont {Shang},
  \citenamefont {Wang}, \citenamefont {Li},\ and\ \citenamefont
  {Lu}}]{shang2018}%
  \BibitemOpen
  \bibfield  {author} {\bibinfo {author} {\bibfnamefont {Y.}~\bibnamefont
  {Shang}}, \bibinfo {author} {\bibfnamefont {Y.}~\bibnamefont {Wang}},
  \bibinfo {author} {\bibfnamefont {M.}~\bibnamefont {Li}},\ and\ \bibinfo
  {author} {\bibfnamefont {R.}~\bibnamefont {Lu}},\ }\href
  {https://doi.org/10.1209/0295-5075/124/60009} {\bibfield  {journal} {\bibinfo
   {journal} {Europhysics Letters}\ }\textbf {\bibinfo {volume} {124}},\
  \bibinfo {pages} {60009} (\bibinfo {year} {2019})}\BibitemShut {NoStop}%
\bibitem [{\citenamefont {Mackay}\ \emph {et~al.}(2002)\citenamefont {Mackay},
  \citenamefont {Bartlett}, \citenamefont {Stephenson},\ and\ \citenamefont
  {Sanders}}]{mackay2002}%
  \BibitemOpen
  \bibfield  {author} {\bibinfo {author} {\bibfnamefont {T.~D.}\ \bibnamefont
  {Mackay}}, \bibinfo {author} {\bibfnamefont {S.~D.}\ \bibnamefont
  {Bartlett}}, \bibinfo {author} {\bibfnamefont {L.~T.}\ \bibnamefont
  {Stephenson}},\ and\ \bibinfo {author} {\bibfnamefont {B.~C.}\ \bibnamefont
  {Sanders}},\ }\href {https://doi.org/10.1088/0305-4470/35/12/304} {\bibfield
  {journal} {\bibinfo  {journal} {Journal of Physics A: Mathematical and
  General}\ }\textbf {\bibinfo {volume} {35}},\ \bibinfo {pages} {2745}
  (\bibinfo {year} {2002})}\BibitemShut {NoStop}%
\bibitem [{\citenamefont {Hamilton}\ \emph {et~al.}(2011)\citenamefont
  {Hamilton}, \citenamefont {Gábris}, \citenamefont {Jex},\ and\ \citenamefont
  {Barnett}}]{Hamilton_2011}%
  \BibitemOpen
  \bibfield  {author} {\bibinfo {author} {\bibfnamefont {C.~S.}\ \bibnamefont
  {Hamilton}}, \bibinfo {author} {\bibfnamefont {A.}~\bibnamefont {Gábris}},
  \bibinfo {author} {\bibfnamefont {I.}~\bibnamefont {Jex}},\ and\ \bibinfo
  {author} {\bibfnamefont {S.~M.}\ \bibnamefont {Barnett}},\ }\href
  {https://doi.org/10.1088/1367-2630/13/1/013015} {\bibfield  {journal}
  {\bibinfo  {journal} {New Journal of Physics}\ }\textbf {\bibinfo {volume}
  {13}},\ \bibinfo {pages} {013015} (\bibinfo {year} {2011})}\BibitemShut
  {NoStop}%
\bibitem [{\citenamefont {Falkner}\ and\ \citenamefont
  {Boettcher}(2014)}]{PhysRevA.90.012307}%
  \BibitemOpen
  \bibfield  {author} {\bibinfo {author} {\bibfnamefont {S.}~\bibnamefont
  {Falkner}}\ and\ \bibinfo {author} {\bibfnamefont {S.}~\bibnamefont
  {Boettcher}},\ }\href {https://doi.org/10.1103/PhysRevA.90.012307} {\bibfield
   {journal} {\bibinfo  {journal} {Phys. Rev. A}\ }\textbf {\bibinfo {volume}
  {90}},\ \bibinfo {pages} {012307} (\bibinfo {year} {2014})}\BibitemShut
  {NoStop}%
\bibitem [{\citenamefont {Lorz}\ \emph {et~al.}(2019)\citenamefont {Lorz},
  \citenamefont {Meyer-Scott}, \citenamefont {Nitsche}, \citenamefont
  {Poto\ifmmode~\check{c}\else \v{c}\fi{}ek}, \citenamefont {G\'abris},
  \citenamefont {Barkhofen}, \citenamefont {Jex},\ and\ \citenamefont
  {Silberhorn}}]{lorz2019photonic}%
  \BibitemOpen
  \bibfield  {author} {\bibinfo {author} {\bibfnamefont {L.}~\bibnamefont
  {Lorz}}, \bibinfo {author} {\bibfnamefont {E.}~\bibnamefont {Meyer-Scott}},
  \bibinfo {author} {\bibfnamefont {T.}~\bibnamefont {Nitsche}}, \bibinfo
  {author} {\bibfnamefont {V.}~\bibnamefont {Poto\ifmmode~\check{c}\else
  \v{c}\fi{}ek}}, \bibinfo {author} {\bibfnamefont {A.}~\bibnamefont
  {G\'abris}}, \bibinfo {author} {\bibfnamefont {S.}~\bibnamefont {Barkhofen}},
  \bibinfo {author} {\bibfnamefont {I.}~\bibnamefont {Jex}},\ and\ \bibinfo
  {author} {\bibfnamefont {C.}~\bibnamefont {Silberhorn}},\ }\href
  {https://doi.org/10.1103/PhysRevResearch.1.033036} {\bibfield  {journal}
  {\bibinfo  {journal} {Phys. Rev. Res.}\ }\textbf {\bibinfo {volume} {1}},\
  \bibinfo {pages} {033036} (\bibinfo {year} {2019})}\BibitemShut {NoStop}%
\bibitem [{\citenamefont {{\v{S}}tefa{\v{n}}{\'a}k}\ \emph
  {et~al.}(2012)\citenamefont {{\v{S}}tefa{\v{n}}{\'a}k}, \citenamefont
  {Bezd{\v{e}}kov{\'a}},\ and\ \citenamefont {Jex}}]{stefanak2012continuous}%
  \BibitemOpen
  \bibfield  {author} {\bibinfo {author} {\bibfnamefont {M.}~\bibnamefont
  {{\v{S}}tefa{\v{n}}{\'a}k}}, \bibinfo {author} {\bibfnamefont
  {I.}~\bibnamefont {Bezd{\v{e}}kov{\'a}}},\ and\ \bibinfo {author}
  {\bibfnamefont {I.}~\bibnamefont {Jex}},\ }\href
  {https://doi.org/10.1140/epjd/e2012-30146-9} {\bibfield  {journal} {\bibinfo
  {journal} {The European Physical Journal D}\ }\textbf {\bibinfo {volume}
  {66}},\ \bibinfo {pages} {142} (\bibinfo {year} {2012})}\BibitemShut
  {NoStop}%
\bibitem [{\citenamefont {Moreva}\ \emph {et~al.}(2006)\citenamefont {Moreva},
  \citenamefont {Maslennikov}, \citenamefont {Straupe},\ and\ \citenamefont
  {Kulik}}]{PhysRevLett.97.023602}%
  \BibitemOpen
  \bibfield  {author} {\bibinfo {author} {\bibfnamefont {E.~V.}\ \bibnamefont
  {Moreva}}, \bibinfo {author} {\bibfnamefont {G.~A.}\ \bibnamefont
  {Maslennikov}}, \bibinfo {author} {\bibfnamefont {S.~S.}\ \bibnamefont
  {Straupe}},\ and\ \bibinfo {author} {\bibfnamefont {S.~P.}\ \bibnamefont
  {Kulik}},\ }\href {https://doi.org/10.1103/PhysRevLett.97.023602} {\bibfield
  {journal} {\bibinfo  {journal} {Phys. Rev. Lett.}\ }\textbf {\bibinfo
  {volume} {97}},\ \bibinfo {pages} {023602} (\bibinfo {year}
  {2006})}\BibitemShut {NoStop}%
\bibitem [{\citenamefont {Annabestani}\ \emph {et~al.}(2022)\citenamefont
  {Annabestani}, \citenamefont {Hassani}, \citenamefont {Tamascelli},\ and\
  \citenamefont {Paris}}]{annabestani2022multiparameter}%
  \BibitemOpen
  \bibfield  {author} {\bibinfo {author} {\bibfnamefont {M.}~\bibnamefont
  {Annabestani}}, \bibinfo {author} {\bibfnamefont {M.}~\bibnamefont
  {Hassani}}, \bibinfo {author} {\bibfnamefont {D.}~\bibnamefont
  {Tamascelli}},\ and\ \bibinfo {author} {\bibfnamefont {M.~G.~A.}\
  \bibnamefont {Paris}},\ }\href {https://doi.org/10.1103/PhysRevA.105.062411}
  {\bibfield  {journal} {\bibinfo  {journal} {Phys. Rev. A}\ }\textbf {\bibinfo
  {volume} {105}},\ \bibinfo {pages} {062411} (\bibinfo {year}
  {2022})}\BibitemShut {NoStop}%
\bibitem [{\citenamefont {Singh}\ \emph {et~al.}(2019)\citenamefont {Singh},
  \citenamefont {Chandrashekar},\ and\ \citenamefont
  {Paris}}]{singh2019quantum}%
  \BibitemOpen
  \bibfield  {author} {\bibinfo {author} {\bibfnamefont {S.}~\bibnamefont
  {Singh}}, \bibinfo {author} {\bibfnamefont {C.~M.}\ \bibnamefont
  {Chandrashekar}},\ and\ \bibinfo {author} {\bibfnamefont {M.~G.~A.}\
  \bibnamefont {Paris}},\ }\href {https://doi.org/10.1103/PhysRevA.99.052117}
  {\bibfield  {journal} {\bibinfo  {journal} {Phys. Rev. A}\ }\textbf {\bibinfo
  {volume} {99}},\ \bibinfo {pages} {052117} (\bibinfo {year}
  {2019})}\BibitemShut {NoStop}%
\bibitem [{\citenamefont {Chandrashekar}\ \emph {et~al.}(2008)\citenamefont
  {Chandrashekar}, \citenamefont {Srikanth},\ and\ \citenamefont
  {Laflamme}}]{PhysRevA.77.032326}%
  \BibitemOpen
  \bibfield  {author} {\bibinfo {author} {\bibfnamefont {C.~M.}\ \bibnamefont
  {Chandrashekar}}, \bibinfo {author} {\bibfnamefont {R.}~\bibnamefont
  {Srikanth}},\ and\ \bibinfo {author} {\bibfnamefont {R.}~\bibnamefont
  {Laflamme}},\ }\href {https://doi.org/10.1103/PhysRevA.77.032326} {\bibfield
  {journal} {\bibinfo  {journal} {Phys. Rev. A}\ }\textbf {\bibinfo {volume}
  {77}},\ \bibinfo {pages} {032326} (\bibinfo {year} {2008})}\BibitemShut
  {NoStop}%
\bibitem [{\citenamefont {Lee}\ \emph {et~al.}(2011)\citenamefont {Lee},
  \citenamefont {Mittal}, \citenamefont {Reichardt}, \citenamefont {Spalek},\
  and\ \citenamefont {Szegedy}}]{inproceedings}%
  \BibitemOpen
  \bibfield  {author} {\bibinfo {author} {\bibfnamefont {T.}~\bibnamefont
  {Lee}}, \bibinfo {author} {\bibfnamefont {R.}~\bibnamefont {Mittal}},
  \bibinfo {author} {\bibfnamefont {B.}~\bibnamefont {Reichardt}}, \bibinfo
  {author} {\bibfnamefont {R.}~\bibnamefont {Spalek}},\ and\ \bibinfo {author}
  {\bibfnamefont {M.}~\bibnamefont {Szegedy}}\ }(\bibinfo {year} {2011})\ pp.\
  \bibinfo {pages} {344 -- 353}\BibitemShut {NoStop}%
\bibitem [{\citenamefont {Chawla}\ \emph {et~al.}(2023)\citenamefont {Chawla},
  \citenamefont {Singh}, \citenamefont {Agarwal}, \citenamefont {Srinivasan},\
  and\ \citenamefont {Chandrashekar}}]{Chawla2023}%
  \BibitemOpen
  \bibfield  {author} {\bibinfo {author} {\bibfnamefont {P.}~\bibnamefont
  {Chawla}}, \bibinfo {author} {\bibfnamefont {S.}~\bibnamefont {Singh}},
  \bibinfo {author} {\bibfnamefont {A.}~\bibnamefont {Agarwal}}, \bibinfo
  {author} {\bibfnamefont {S.}~\bibnamefont {Srinivasan}},\ and\ \bibinfo
  {author} {\bibfnamefont {C.~M.}\ \bibnamefont {Chandrashekar}},\ }\href
  {https://doi.org/10.1038/s41598-023-39061-1} {\bibfield  {journal} {\bibinfo
  {journal} {Scientific Reports}\ }\textbf {\bibinfo {volume} {13}} (\bibinfo
  {year} {2023})}\BibitemShut {NoStop}%
\bibitem [{\citenamefont {Wang}\ \emph {et~al.}(2021)\citenamefont {Wang},
  \citenamefont {Lu}, \citenamefont {Zhang},\ and\ \citenamefont
  {Liu}}]{wang2021qsim}%
  \BibitemOpen
  \bibfield  {author} {\bibinfo {author} {\bibfnamefont {X.}~\bibnamefont
  {Wang}}, \bibinfo {author} {\bibfnamefont {K.}~\bibnamefont {Lu}}, \bibinfo
  {author} {\bibfnamefont {Y.}~\bibnamefont {Zhang}},\ and\ \bibinfo {author}
  {\bibfnamefont {K.}~\bibnamefont {Liu}},\ }\href@noop {} {\bibfield
  {journal} {\bibinfo  {journal} {Applied Intelligence}\ }\textbf {\bibinfo
  {volume} {51}},\ \bibinfo {pages} {2574} (\bibinfo {year}
  {2021})}\BibitemShut {NoStop}%
\bibitem [{\citenamefont {Zatelli}\ \emph {et~al.}(2020)\citenamefont
  {Zatelli}, \citenamefont {Benedetti},\ and\ \citenamefont
  {Paris}}]{zatelli2020scattering}%
  \BibitemOpen
  \bibfield  {author} {\bibinfo {author} {\bibfnamefont {F.}~\bibnamefont
  {Zatelli}}, \bibinfo {author} {\bibfnamefont {C.}~\bibnamefont {Benedetti}},\
  and\ \bibinfo {author} {\bibfnamefont {M.~G.}\ \bibnamefont {Paris}},\
  }\href@noop {} {\bibfield  {journal} {\bibinfo  {journal} {Entropy}\ }\textbf
  {\bibinfo {volume} {22}},\ \bibinfo {pages} {1321} (\bibinfo {year}
  {2020})}\BibitemShut {NoStop}%
\bibitem [{\citenamefont {Mallick}\ \emph {et~al.}(2020)\citenamefont
  {Mallick}, \citenamefont {Fistul}, \citenamefont {Kaczynska},\ and\
  \citenamefont {Flach}}]{mallick2020spectral}%
  \BibitemOpen
  \bibfield  {author} {\bibinfo {author} {\bibfnamefont {A.}~\bibnamefont
  {Mallick}}, \bibinfo {author} {\bibfnamefont {M.}~\bibnamefont {Fistul}},
  \bibinfo {author} {\bibfnamefont {P.}~\bibnamefont {Kaczynska}},\ and\
  \bibinfo {author} {\bibfnamefont {S.}~\bibnamefont {Flach}},\ }\href@noop {}
  {\bibfield  {journal} {\bibinfo  {journal} {Physical Review A}\ }\textbf
  {\bibinfo {volume} {101}},\ \bibinfo {pages} {032119} (\bibinfo {year}
  {2020})}\BibitemShut {NoStop}%
\bibitem [{Note1()}]{Note1}%
  \BibitemOpen
  \bibinfo {note} {I.e., a coin operator which does depend neither on time nor
  on the position of the walker.}\BibitemShut {Stop}%
\bibitem [{\citenamefont {Cacciatori}\ and\ \citenamefont
  {Scotti}(2022)}]{cacciatori2022compact}%
  \BibitemOpen
  \bibfield  {author} {\bibinfo {author} {\bibfnamefont {S.~L.}\ \bibnamefont
  {Cacciatori}}\ and\ \bibinfo {author} {\bibfnamefont {A.}~\bibnamefont
  {Scotti}},\ }\href@noop {} {\bibfield  {journal} {\bibinfo  {journal}
  {Universe}\ }\textbf {\bibinfo {volume} {8}},\ \bibinfo {pages} {492}
  (\bibinfo {year} {2022})}\BibitemShut {NoStop}%
\bibitem [{\citenamefont {Byrd}(1997)}]{byrd1997geometry}%
  \BibitemOpen
  \bibfield  {author} {\bibinfo {author} {\bibfnamefont {M.}~\bibnamefont
  {Byrd}},\ }\href@noop {} {\bibfield  {journal} {\bibinfo  {journal} {arXiv
  preprint physics/9708015}\ } (\bibinfo {year} {1997})}\BibitemShut {NoStop}%
\bibitem [{\citenamefont {Endo}\ \emph {et~al.}(2020)\citenamefont {Endo},
  \citenamefont {Endo}, \citenamefont {Komatsu},\ and\ \citenamefont
  {Konno}}]{endo2020eigenvalues}%
  \BibitemOpen
  \bibfield  {author} {\bibinfo {author} {\bibfnamefont {S.}~\bibnamefont
  {Endo}}, \bibinfo {author} {\bibfnamefont {T.}~\bibnamefont {Endo}}, \bibinfo
  {author} {\bibfnamefont {T.}~\bibnamefont {Komatsu}},\ and\ \bibinfo {author}
  {\bibfnamefont {N.}~\bibnamefont {Konno}},\ }\href@noop {} {\bibfield
  {journal} {\bibinfo  {journal} {Entropy}\ }\textbf {\bibinfo {volume} {22}},\
  \bibinfo {pages} {127} (\bibinfo {year} {2020})}\BibitemShut {NoStop}%
\bibitem [{\citenamefont {Li}\ \emph {et~al.}(2015)\citenamefont {Li},
  \citenamefont {Mc~Gettrick}, \citenamefont {Zhang},\ and\ \citenamefont
  {Zhang}}]{li2015one}%
  \BibitemOpen
  \bibfield  {author} {\bibinfo {author} {\bibfnamefont {D.}~\bibnamefont
  {Li}}, \bibinfo {author} {\bibfnamefont {M.}~\bibnamefont {Mc~Gettrick}},
  \bibinfo {author} {\bibfnamefont {W.-W.}\ \bibnamefont {Zhang}},\ and\
  \bibinfo {author} {\bibfnamefont {K.-J.}\ \bibnamefont {Zhang}},\ }\href@noop
  {} {\bibfield  {journal} {\bibinfo  {journal} {Chinese Physics B}\ }\textbf
  {\bibinfo {volume} {24}},\ \bibinfo {pages} {050305} (\bibinfo {year}
  {2015})}\BibitemShut {NoStop}%
\bibitem [{\citenamefont {Moore}\ and\ \citenamefont
  {Russell}(2002)}]{moore2002proceedings}%
  \BibitemOpen
  \bibfield  {author} {\bibinfo {author} {\bibfnamefont {C.}~\bibnamefont
  {Moore}}\ and\ \bibinfo {author} {\bibfnamefont {A.}~\bibnamefont
  {Russell}},\ }in\ \href@noop {} {\emph {\bibinfo {booktitle} {Proceedings of
  the 6th International Workshop on Randomization and Approximation Techniques
  in Computer Science (RANDOM’02)}}},\ \bibinfo {editor} {edited by\ \bibinfo
  {editor} {\bibfnamefont {J.~D.~P.}\ \bibnamefont {Rolim}}\ and\ \bibinfo
  {editor} {\bibfnamefont {P.}~\bibnamefont {Vadham}}}\ (\bibinfo  {publisher}
  {Springer, Cambridge, MA},\ \bibinfo {year} {2002})\ p.\ \bibinfo {pages}
  {164–178}\BibitemShut {NoStop}%
\bibitem [{\citenamefont {Tregenna}\ \emph {et~al.}(2003)\citenamefont
  {Tregenna}, \citenamefont {Flanagan}, \citenamefont {Maile},\ and\
  \citenamefont {Kendon}}]{tregenna2003controlling}%
  \BibitemOpen
  \bibfield  {author} {\bibinfo {author} {\bibfnamefont {B.}~\bibnamefont
  {Tregenna}}, \bibinfo {author} {\bibfnamefont {W.}~\bibnamefont {Flanagan}},
  \bibinfo {author} {\bibfnamefont {R.}~\bibnamefont {Maile}},\ and\ \bibinfo
  {author} {\bibfnamefont {V.}~\bibnamefont {Kendon}},\ }\href@noop {}
  {\bibfield  {journal} {\bibinfo  {journal} {New Journal of Physics}\ }\textbf
  {\bibinfo {volume} {5}},\ \bibinfo {pages} {83} (\bibinfo {year}
  {2003})}\BibitemShut {NoStop}%
\bibitem [{\citenamefont {{\v{S}}tefa{\v{n}}{\'a}k}\ \emph
  {et~al.}(2014)\citenamefont {{\v{S}}tefa{\v{n}}{\'a}k}, \citenamefont
  {Bezd{\v{e}}kov{\'a}},\ and\ \citenamefont {Jex}}]{stefanak2014limit}%
  \BibitemOpen
  \bibfield  {author} {\bibinfo {author} {\bibfnamefont {M.}~\bibnamefont
  {{\v{S}}tefa{\v{n}}{\'a}k}}, \bibinfo {author} {\bibfnamefont
  {I.}~\bibnamefont {Bezd{\v{e}}kov{\'a}}},\ and\ \bibinfo {author}
  {\bibfnamefont {I.}~\bibnamefont {Jex}},\ }\href@noop {} {\bibfield
  {journal} {\bibinfo  {journal} {Physical Review A}\ }\textbf {\bibinfo
  {volume} {90}},\ \bibinfo {pages} {012342} (\bibinfo {year}
  {2014})}\BibitemShut {NoStop}%
\bibitem [{\citenamefont {Sarkar}\ \emph {et~al.}(2020)\citenamefont {Sarkar},
  \citenamefont {Mandal},\ and\ \citenamefont
  {Adhikari}}]{sarkar2020periodicity}%
  \BibitemOpen
  \bibfield  {author} {\bibinfo {author} {\bibfnamefont {R.~S.}\ \bibnamefont
  {Sarkar}}, \bibinfo {author} {\bibfnamefont {A.}~\bibnamefont {Mandal}},\
  and\ \bibinfo {author} {\bibfnamefont {B.}~\bibnamefont {Adhikari}},\
  }\href@noop {} {\bibfield  {journal} {\bibinfo  {journal} {Linear Algebra and
  its Applications}\ }\textbf {\bibinfo {volume} {604}},\ \bibinfo {pages}
  {399} (\bibinfo {year} {2020})}\BibitemShut {NoStop}%
\bibitem [{\citenamefont {Mandal}\ \emph {et~al.}(2022)\citenamefont {Mandal},
  \citenamefont {Sarkar}, \citenamefont {Chakraborty},\ and\ \citenamefont
  {Adhikari}}]{mandal2022limit}%
  \BibitemOpen
  \bibfield  {author} {\bibinfo {author} {\bibfnamefont {A.}~\bibnamefont
  {Mandal}}, \bibinfo {author} {\bibfnamefont {R.~S.}\ \bibnamefont {Sarkar}},
  \bibinfo {author} {\bibfnamefont {S.}~\bibnamefont {Chakraborty}},\ and\
  \bibinfo {author} {\bibfnamefont {B.}~\bibnamefont {Adhikari}},\ }\href
  {https://doi.org/10.1103/PhysRevA.106.042405} {\bibfield  {journal} {\bibinfo
   {journal} {Phys. Rev. A}\ }\textbf {\bibinfo {volume} {106}},\ \bibinfo
  {pages} {042405} (\bibinfo {year} {2022})}\BibitemShut {NoStop}%
\bibitem [{Note2()}]{Note2}%
  \BibitemOpen
  \bibinfo {note} {All the points are equivalent in the infinite line, so our
  assumption is just to have an initially localized walker.}\BibitemShut
  {Stop}%
\bibitem [{\citenamefont {Mendaš}(2008)}]{mendas2008}%
  \BibitemOpen
  \bibfield  {author} {\bibinfo {author} {\bibfnamefont {I.~P.}\ \bibnamefont
  {Mendaš}},\ }\href {https://doi.org/10.1063/1.2982276} {\bibfield  {journal}
  {\bibinfo  {journal} {Journal of Mathematical Physics}\ }\textbf {\bibinfo
  {volume} {49}},\ \bibinfo {pages} {092102} (\bibinfo {year}
  {2008})}\BibitemShut {NoStop}%
\bibitem [{\citenamefont {Caves}\ and\ \citenamefont
  {Milburn}(2000)}]{CAVES2000439}%
  \BibitemOpen
  \bibfield  {author} {\bibinfo {author} {\bibfnamefont {C.~M.}\ \bibnamefont
  {Caves}}\ and\ \bibinfo {author} {\bibfnamefont {G.~J.}\ \bibnamefont
  {Milburn}},\ }\href
  {https://doi.org/https://doi.org/10.1016/S0030-4018(99)00693-8} {\bibfield
  {journal} {\bibinfo  {journal} {Optics Communications}\ }\textbf {\bibinfo
  {volume} {179}},\ \bibinfo {pages} {439} (\bibinfo {year}
  {2000})}\BibitemShut {NoStop}%
\bibitem [{\citenamefont {Paris}(2009)}]{paris2009quantum}%
  \BibitemOpen
  \bibfield  {author} {\bibinfo {author} {\bibfnamefont {M.~G.}\ \bibnamefont
  {Paris}},\ }\href@noop {} {\bibfield  {journal} {\bibinfo  {journal}
  {International Journal of Quantum Information}\ }\textbf {\bibinfo {volume}
  {7}},\ \bibinfo {pages} {125} (\bibinfo {year} {2009})}\BibitemShut {NoStop}%
\bibitem [{\citenamefont {Curtright}\ \emph {et~al.}(2014)\citenamefont
  {Curtright}, \citenamefont {Fairlie}, \citenamefont {Zachos} \emph
  {et~al.}}]{curtright2014compact}%
  \BibitemOpen
  \bibfield  {author} {\bibinfo {author} {\bibfnamefont {T.~L.}\ \bibnamefont
  {Curtright}}, \bibinfo {author} {\bibfnamefont {D.~B.}\ \bibnamefont
  {Fairlie}}, \bibinfo {author} {\bibfnamefont {C.~K.}\ \bibnamefont {Zachos}},
  \emph {et~al.},\ }\href@noop {} {\bibfield  {journal} {\bibinfo  {journal}
  {SIGMA. Symmetry, Integrability and Geometry: Methods and Applications}\
  }\textbf {\bibinfo {volume} {10}},\ \bibinfo {pages} {084} (\bibinfo {year}
  {2014})}\BibitemShut {NoStop}%
\bibitem [{\citenamefont {Parthasarathy}(2001)}]{parthasarathy2001consistency}%
  \BibitemOpen
  \bibfield  {author} {\bibinfo {author} {\bibfnamefont {K.}~\bibnamefont
  {Parthasarathy}},\ }in\ \href@noop {} {\emph {\bibinfo {booktitle}
  {Stochastics in finite and infinite dimensions}}},\ \bibinfo {editor} {edited
  by\ \bibinfo {editor} {\bibfnamefont {T.}~\bibnamefont {Hida}}, \bibinfo
  {editor} {\bibfnamefont {R.~L.}\ \bibnamefont {Karandikar}}, \bibinfo
  {editor} {\bibfnamefont {H.}~\bibnamefont {Kunita}}, \bibinfo {editor}
  {\bibfnamefont {B.~S.}\ \bibnamefont {Rajput}}, \bibinfo {editor}
  {\bibfnamefont {S.}~\bibnamefont {Watanabe}},\ and\ \bibinfo {editor}
  {\bibfnamefont {J.}~\bibnamefont {Xiong}}}\ (\bibinfo  {publisher}
  {Springer},\ \bibinfo {address} {New York},\ \bibinfo {year} {2001})\ pp.\
  \bibinfo {pages} {361--377}\BibitemShut {NoStop}%
\end{thebibliography}%

\end{document}